\documentclass[11pt]{article}
\usepackage{amssymb,latexsym, amsmath}
\usepackage{graphicx}
\makeatletter
\@addtoreset{equation}{section}

\makeatother

\newtheorem{theorem}{Theorem}[section]

\newtheorem{rem}[theorem]{Remark}

\newtheorem{exa}[theorem]{Example}

\newtheorem{ques}[theorem]{Question}

\def\mathbb#1{\mathbf{ #1}}
\def\mathfrak#1{\mathbf{ #1}}
\def\bye { \end{document} }
\def\refce{\smallskip \noindent\hangindent\parindent}
\def\frak#1#2{\textstyle\frac{#1}{#2}}
\def\half{\textstyle\frac{1}{2}}

\newcount\Figsinline \def\Figsinline{1}

\begin{document}

\title{
%
{\it Stepwise Precession of the
Resonant Swinging Spring}
}
\author{
Darryl D. Holm\footnote{ email: dholm@lanl.gov}
\\Theoretical Division and CNLS
\\Los Alamos National
Laboratory, MS B284
\\ Los Alamos, NM 87545
\and
Peter Lynch\footnote{ email: Peter.Lynch@met.ie}
\\Met \'Eireann
\\Glasnevin Hill
\\ Dublin 9, Ireland
}

\date{
}

\maketitle

\large

\begin{abstract}
The swinging spring, or elastic pendulum, has a 2:1:1 resonance arising at
cubic
order in its approximate Lagrangian. The corresponding modulation
equations are the well-known three-wave equations that also apply, for example,
in laser-matter interaction in a cavity. We use Hamiltonian reduction and
pattern evocation techniques to derive a formula that describes the
characteristic feature of this system's dynamics, namely, the stepwise
precession of its azimuthal angle.
\smallskip

\noindent
PACS numbers:   02.40.-k, 05.45.-a, 45.10.Db, 45.20.Jj
\smallskip

\noindent
Keywords: Classical mechanics, Variational principles, Averaged Lagrangian,
Elastic Pendulum, Nonlinear Resonance.

\end{abstract}

\clearpage

\section{Introduction}

\subsection{Problem statement, approach and summary of results}
The elastic pendulum or swinging spring is a simple mechanical system
that exhibits complex dynamics. It consists of a heavy mass suspended
from a fixed point by a light spring which can stretch but not bend,
moving under gravity. We investigate the 2:1:1 resonance dynamics of this
system in three dimensions and study its characteristic feature -- the
regular stepwise precession of its azimuthal angle.

When the Lagrangian is approximated to cubic order and averaged over the fast
dynamics, the resulting modulation equations have three independent
constants of
motion and are completely integrable.  These modulation equations are
identical to the three-wave equations for resonant triad interactions in fluids
and plasmas, and in laser-matter interaction.  We reduce the system to a form
amenable to analytical solution and show how the full solution may be
reconstructed.  We examine the geometry of the solutions in phase-space and
develop a number of simple qualitative descriptions of the motion.

We compare solutions of the exact and modulation equations and show
that they are remarkably similar.  A characteristic stepwise precession occurs
as the motion cycles between quasi-vertical and quasi-horizontal. That is,
during each quasi-vertical phase, the azimuth of the swing plane precesses
by a constant angular increment. This stepwise azimuthal precession
occurs in bursts, when the motion is nearly vertical. By transforming to
non-uniformly rotating coordinates and assuming a geometric constraint
(essentially the method of pattern evocation), we find a formula for the
rotation of the swing plane.  This formula gives a highly accurate description
of the stepwise precession of the azimuthal angle of the motion.

\subsection{History of the problem}
The first comprehensive analysis of the elastic pendulum appeared in
Vitt and Gorelik (1933). These authors were inspired by the analogy
between this system and the Fermi resonance of a carbon-dioxide
molecule.  We make connections in this paper with other physical
systems of current interest.
For example, we show that the modulation equations for the averaged
motion of the swinging spring may be transformed into the equations for
three-wave interactions that appear in analyzing fluid and plasma
systems, and in laser-matter interaction.  These three complex equations are
also identical to the Maxwell-Schr\"odinger envelope equations for the
interaction between radiation and a two-level resonant medium in a microwave
cavity (Holm and Kova\v ci\v c, 1992).
The three-wave equations also govern the envelope dynamics of light
waves in an inhomogeneous material (David {\it et al.}, 1990, Alber
{\it et al.}, 1998a,b). For the special case where the Hamiltonian takes
the value zero, the equations reduce to Euler's equations for a freely
rotating rigid body. Finally, the equations are also equivalent to a
complex (unforced and undamped) version of the Lorenz (1963)
three-component  model, which has been the subject of many studies
(Sparrow, 1982).  Thus, the simple spring pendulum, which was first
studied to provide a classical analogue to the quantum phenomenon of
Fermi resonance, now provides a concrete mechanical system
which simulates a wide range of physical phenomena.

All of the previous studies of the spring pendulum known to us have
considered motion in two dimensions. To our knowledge, only Cayton
(1977) discussed three-dimensional solutions and observed the curious
rotation of the swing plane between successive cycles when the
horizontal energy is maximum.  This particular aspect of the behavior
of the swinging spring in three dimensions is its most striking
difference from two-dimensional motions.  Suppose the system is excited
initially near its vertical oscillation mode.  Since purely vertical
motion is unstable, horizontal motion soon develops. The horizontal
oscillations grow to a maximum and then subside again.
An alternating cycle of quasi-vertical and quasi-horizontal oscillations
recurs indefinitely.  Seen from above, during each
horizontal excursion of several oscillations the projected motion is
approximately elliptical.  Experimentally and numerically one observes
that between any two successive horizontal excursions the orientation
of the projected ellipse rotates by the same angle, thereby causing a
{\it stepwise precession of the swing plane}. In principle, the precession
angle between successive horizontal excursions can be deduced from the
complete solution of the integrable envelope equations. We seek a
simple approximate expression for the precession of the swing plane in
terms of the solution of the reduced dynamics.

Lynch (2001) found a particular solution for the rate of precession of
the swing plane by using the method of multiple time scales in rotating
coordinates and introducing a certain angular solution Ansatz.  We
recover Lynch's particular solution among a family of other solutions
for the swing plane precession rate.  This family is obtained via the
method of averaged Lagrangians by seeking solutions of the modulation
equations that satisfy a geometrical constraint of being ``instantaneously
elliptical.''  We apply the method of {\it pattern evocation in shape space}
(Marsden, {\it et al.}, 1995, 1996). Using this process, one identifies
patterns by viewing the dynamics relative to rotating frames with certain
critical angular velocities.  Our numerical integrations show that the
solution resulting from this geometrical postulate estimates the precession of
the swing plane with surprisingly high accuracy.


\section{Equations of motion}

The physical system under investigation is an elastic pendulum, or
swinging spring, consisting of a heavy mass suspended from a fixed
point by a light spring  which can stretch but not bend, moving under
gravity, $g$. We assume an unstretched length $\ell_0$, length $\ell$ at
equilibrium, spring constant $k$ and unit mass $m=1$.
The corresponding Lagrangian, approximated to cubic order in the amplitudes, is
\begin{equation}\label{Lag-3D-0}
L = \half \left( \dot x^2+\dot y^2+\dot z^2 \right)
 -  \half \left( \omega_R^2(x^2+y^2)+\omega_Z^2 z^2 \right)
 +  \half \lambda(x^2+y^2) z \, ,
\end{equation}
where $x$, $y$ and $z$ are Cartesian coordinates centered at the point
of equilibrium, $\omega_R=\sqrt{g/\ell}$ is the frequency of linear
pendular motion, $\omega_Z=\sqrt{k/m}$ is the frequency of its elastic
oscillations and $\lambda=\ell_0\omega_Z^2/\ell^2$.
The system is illustrated schematically in Fig.~1.
The Euler-Lagrange equations of motion may be written
\begin{eqnarray}
\ddot x + \omega_R^2 x &=& \lambda xz              \label{XXeqn}  \,,\\
\ddot y + \omega_R^2 y &=& \lambda yz              \label{YYeqn}  \,,\\
\ddot z + \omega_Z^2 z &=& \half\lambda (x^2+y^2)  \label{ZZeqn}  \,.
\end{eqnarray}
There are two constants of the motion, the total energy $E$
and the angular momentum $h$ given by
\begin{eqnarray*}
E
&=&
\half \left( \dot x^2+\dot y^2+\dot z^2 \right)
 +  \half \left( \omega_R^2(x^2+y^2)+\omega_Z^2 z^2 \right)
 -  \half \lambda(x^2+y^2) z
\,,\quad
h
=
( x \dot y - y \dot x ) \,.
\end{eqnarray*}
The system is not integrable.
Its chaotic motions have been studied by many authors (see, e.g.,
Refs. in Lynch, 2001). Previous studies have considered the two
dimensional case, for which the angular momentum $h$ vanishes.


We confine attention to the resonant case $\omega_Z=2\omega_R$
and apply the averaged Lagrangian technique (Whitham, 1974).
The solution of (\ref{XXeqn})--(\ref{ZZeqn}) is assumed to be of the form
\begin{eqnarray}
x &=& \Re[a_0(t)\exp(i\omega_R t)] \, ,  \label{ModSolA} \\
y &=& \Re[b_0(t)\exp(i\omega_R t)] \, ,  \label{ModSolB} \\
z &=& \Re[c_0(t)\exp(2i\omega_R t)] \,.  \label{ModSolC}
\end{eqnarray}
(The zero-subscripts in $a_0$, $b_0$ and $c_0$ are introduced to
distinguish from the variables $a$, $b$ and $c$ in a rotating frame,
introduced below.) The coefficients $a_0(t)$, $b_0(t)$ and $c_0(t)$ are assumed
to vary on a time scale which is much longer than the time-scale of the
oscillations, $\tau = 1/\omega_R$.
The Lagrangian is averaged over time $\tau$ to give,
$$
\langle L \rangle = \half\omega_R
\Big[  \Im\{\dot a_0 a_0^* + \dot b_0 b_0^* + 2\dot c_0 c_0^*\}
 + \kappa\,\Re\{(a_0^2+b_0^2)c_0^*\} \Big]
\,,$$
where $\kappa=\lambda/(4\omega_R)$.
We regard the quantities $a_0,b_0,c_0$ as generalized coordinates.
The averaged Lagrangian equations of motion are then
\begin{eqnarray}
i\dot a_0 &=& \kappa a_0^*c_0  \,,      \label{niceone} \\
i\dot b_0 &=& \kappa b_0^*c_0   \,,     \label{nicetwo} \\
i\dot c_0 &=& \textstyle{\frac{1}{4}}\kappa (a_0^2+b_0^2)
\label{nicethree}
\end{eqnarray}
Eqns. (\ref{niceone})--(\ref{nicethree}) are
the complex  versions of Eqns.~(68)--(73) in Lynch (2001),
which were derived using the method of multiple time-scale analysis.

We now absorb the constant $\kappa$ by rescaling the time,
$t\rightarrow \kappa t$, and transform variables
\begin{eqnarray*}
A = \half(a_0+ib_0)\, , \quad B = \half(a_0-ib_0)\, , \quad C = c_0
\,.
\end{eqnarray*}
Consequently, the equations of motion take the form
\begin{eqnarray}
i\dot A &=& B^*C \, ,  \label{TWEa}   \\
i\dot B &=& CA^* \, ,  \label{TWEb}   \\
i\dot C &=& AB   \, .  \label{TWEc}
\end{eqnarray}
These three complex equations are well-known as the
{\it three-wave interaction equations},
which govern quadratic wave resonance in fluids and plasmas.

The three-wave interaction equations (\ref{TWEa})-(\ref{TWEc}) may be written
in canonical form with Hamiltonian $H=\Re(ABC^*)$ and Poisson brackets
$\{A,A^*\}=\{B,B^*\}=\{C,C^*\}=-2i$, as
\begin{eqnarray}
i\dot A &=& i\{A,H\} = 2\partial H/ \partial A^* \, ,  \label{TWEaHam}   \\
i\dot B &=& i\{B,H\} = 2\partial H/ \partial B^* \, ,  \label{TWEbHam}   \\
i\dot C &=& i\{C,H\} = 2\partial H/ \partial C^* \, .  \label{TWEcHam}
\end{eqnarray}
These equations conserve the following three quantities,
\begin{eqnarray}
H &=& \half(ABC^*+A^*B^*C)  = \Re(ABC^*)
\,,\label{Hconstant} \\
N &=& |A|^2 + |B|^2 + 2|C|^2
\,,\label{Nconstant} \\
J &=& |A|^2 - |B|^2
\,.\label{Jconstant}
\end{eqnarray}
Thus, the modulation equations for the swinging spring are
transformed into the three-wave equations, which are known to be
completely integrable. See Alber {\it et al.} (1998a) for references
to the three wave equations and an extensive elaboration of their
properties as a paradigm for Hamiltonian reduction.

The following positive-definite combinations of $N$ and $J$ are
physically significant:
$$
N_{+} \equiv \half(N+J) = |A|^2 + |C|^2 \, , \qquad
N_{-} \equiv \half(N-J) = |B|^2 + |C|^2 \, .
$$
These combinations are known as the {\it Manley-Rowe relations} in the
extensive literature about three-wave interactions. The quantities
$H$, $N_{+}$ and $N_{-}$  provide three independent constants of the motion.

\subsection{A brief history of the three-wave equations}\label{3Wave-subsec}

\paragraph{Fluids and plasmas.}
The three-wave equations model the nonlinear dynamics of the amplitudes
of three waves in fluids or plasmas (Bretherton, 1964). The equations
result from a
perturbation analysis of the Charney equation
\begin{equation}
\frac{\partial}{\partial t}[\nabla^2\psi-F\psi] +
\left( \frac{\partial\psi}{\partial x}
\frac{\partial\nabla^2\psi}{\partial y}
-
       \frac{\partial\psi}{\partial y}
\frac{\partial\nabla^2\psi}{\partial x} \right)
+ \beta \frac{\partial\psi}{\partial x} = 0  \label{Charney}
\end{equation}
(see, e.g., Pedlosky, 1987 for theory and notation).
This equation is equivalent to the Hasegawa-Mima
equation describing drift-waves in an inhomogeneous plasma
in a magnetic field (Hasegawa and Mima, 1977).
Longuet-Higgins and Gill (1967) examined the interactions between
planetary Rossby waves in the atmosphere and
derived detailed conditions for three-wave resonance.
The correspondence between Rossby waves in the atmosphere and drift
waves in plasma have been thoroughly explored by Horton and Hasegawa
(1994). Resonant wave-triad interactions play an essential role in the
generation of turbulence and in determining the statistics of the power
spectrum.  Both energy and enstrophy are conserved in fluid systems
governed by the Charney equation (\ref{Charney}).

\paragraph{Laser-matter interaction.}
Equations (\ref{TWEa})--(\ref{TWEc}) are also equivalent to the
Maxwell-Schr\"odinger envelope equations for the interaction between
radiation and a two-level resonant medium in a microwave cavity.
Holm and Kova\v ci\v c (1992) show that perturbations of this system lead to
homoclinic chaos, but we shall not explore that issue here.
Wersinger {\it et al}. (1980) used a forced and damped version
of the three-wave equations to study instability
saturation by nonlinear mode coupling, and found irregular solutions
indicating the presence of a strange attractor. See also Ott (1993) and
Holm, Kova\v ci\v c and Wettergren (1995, 1996) for more detailed studies of
the perturbed three-wave system.


\paragraph{Nonlinear optics.}
The three-wave system also describes the dynamics of the
envelopes of light-waves interacting quadratically in nonlinear
material. The system has been examined in a series of recent papers
(Alber, {\it et al.}, 1998a,b; Luther {\it et al.}, 2000)
using a geometrical approach which allowed the reduced dynamics for the
wave intensities to be represented as motion on a closed surface in
three dimensions  -- the three-wave surface. Information about the
corresponding reconstruction phases was recovered using the theory of
connections on principal bundles.

In the special case $H=0$ the system (\ref{TWEa})--(\ref{TWEc})
reduces  to three real equations.  Let
$$
  A = iX_1\exp(i\phi_1) \,, \quad B = iX_2\exp(i\phi_2) \,,
                            \quad C = iX_3\exp(i(\phi_1+\phi_2))
$$
where $X_1$, $X_2$ and $X_3$ are real and the phases
$\phi_1$ and $\phi_2$ are constants.
The modulation equations become
\begin{equation}
\dot X_1 = -X_2X_3 \,, \quad \dot X_2 = -X_3X_1 \,,
                       \quad \dot X_3 = +X_1X_2 \,.
\label{SpeCaseOne}
\end{equation}
We note that these equations are re-scaled versions of the
Euler equations for the rotation of a free rigid body.
The dynamics in this special case is expressible as motion on
${\bf R}^3$ namely,
\begin{equation}
\dot {\mathbf{X} }
=
\frak{1}{8} \nabla J \times \nabla N
=
\frak{1}{4} \nabla N_{+} \times \nabla N_{-}
\,.
\label{SpeCase}
\end{equation}
Considering the constancy of $J$ and $N$, we can describe a trajectory
of the motion as an
intersection between a hyperbolic cylinder ($J$ constant, see
(\ref{Jconstant}))
and an oblate spheroid ($N$ constant; see (\ref{Nconstant})).
Eqn.~(\ref{SpeCase}) provides an alternative description.
Here we have used the freedom in the ${\bf R}^3$ Poisson
bracket exploited by David and Holm (1992) to represent the equations
of motion on the intersection of two orthogonal circular cylinders,
the level surfaces of the Manley-Rowe quantities, $N_{+}$ and $N_{-}$.
The invariance of the trajectories means that while the level surfaces of
$J$ and $N$ differ from those of $N_{+}$ and $N_{-}$,
their intersections are precisely the same.
For this particular value of $H=0$, the motion may be further reduced by
expressing it in the coordinates lying on one of these two cylinders.
See Holm and Marsden (1991) for the corresponding transformation of rigid body
motion into pendular motion. See David and Holm (1992), and Alber {\it et
al.} (1998a,b) for discussions of geometric phases in this situation.

\subsection{Reduction of the system \& Reconstruction of the solution}

To reduce the system for $H\ne0$, we employ a further canonical
transformation, introduced in Holm and Kova\v ci\v c, 1992. Namely, we set
\begin{eqnarray}
A &=& |A|\exp(i\xi)       \label{TransA}
\,,\\
B &=& |B|\exp(i\eta )     \label{TransB}
\,,\\
C &=& Z \exp(i(\xi+\eta)) \label{TransC}
\, .
\end{eqnarray}
This transformation is canonical -- it preserves the symplectic form
$$
dA\wedge dA^* + dB\wedge dB^* + dC\wedge dC^* = dZ\wedge dZ^* \, .
$$
In these variables, the Hamiltonian is a function of only $Z$  and
$Z^*$
$$
H = \half (Z+Z^*) \cdot \sqrt{ N_{+}-|Z|^2}
                        \cdot \sqrt{ N_{-}-|Z|^2} \, .
$$
The Poisson bracket is $\{Z,Z^*\}=-2i$ and the canonical equations reduce to
$$
i\dot Z = i\{Z,H\} = 2\frac{\partial H }{ \partial Z^*} \, .
$$
This provides the slow dynamics of both the amplitude and phase
of $Z=|Z|e^{i\zeta}$.
\bigskip

The amplitude $|Z|=|C|$ is obtained in closed form in terms of
Jacobi elliptic functions as the solution of
\begin{equation}
\left( \frac{ d{\cal Q}}{ d\tau }\right)^2 =
\big[ {\cal Q}^3-2{\cal Q}^2+(1-{\cal J}^2){\cal Q}+2{\cal E} \big] \, ,
\label{Qeqn}
\end{equation}
where
${\cal Q}=2|Z|^2/N$, ${\cal J}=J/N$, ${\cal E}=-4H^2/N^3$ and
$\tau=\sqrt{2N}t$.
This is equivalent to Eqn.~(75) in Lynch, 2001 (an explicit expression for the
solution in terms of elliptic functions is given in that paper).
Once $|Z|$ is known, $|A|$ and $|B|$ follow immediately from the
Manley-Rowe relations,
$$
|A| = \sqrt{N_{+}-|Z|^2} \,, \qquad |B| = \sqrt{N_{-}-|Z|^2} \, .
$$
The phases $\xi$ and $\eta$ may now be determined. Using the three-wave
equations  (\ref{TWEa})--(\ref{TWEc}) together with
(\ref{TransA})--(\ref{TransC}), one finds
\begin{equation}
\dot\xi = -\frac{ H}{ |A|^2} \, , \qquad \dot\eta = -\frac{ H }{ |B|^2} \, ,
\end{equation}
so that $\xi$ and $\eta$ can be obtained by quadratures.
Finally, the phase $\zeta$ of $Z$ is determined unambiguously by
\begin{equation}
\frac{d|Z|^2}{ dt}
= -2H\tan\zeta
\qquad {\rm and} \qquad H = |A||B||Z|\cos\zeta \,.
\end{equation}
Hence, we can now reconstruct the full solution as,
$$
A=|A|\exp(i\xi) \,, \quad B=|B|\exp(i\eta ) \,,
         \quad C=|Z|\exp\big(i(\xi+\eta+\zeta)\big) \,.
$$


\section{Phase portraits.}

Consider the plane $\cal C$ in phase-space defined by $A=B=0$.
This is a plane of unstable equilibrium points, representing
purely vertical oscillations of the spring.
The Hamiltonian vanishes identically on this plane, as does the angular
momentum $J$.  Each point $c_0$ in $\cal C$ has a heteroclinic
orbit linking it to its antipodal point $-c_0$. Thus, the plane ${\cal C}$ of
critical points is connected to itself by heteroclinic orbits.
In Fig.~2, the horizontal plane is ${\cal C}$ and the vertical plane contains
heteroclinic orbits from $c_0$ to $-c_0$. The vertical axis is
$R=\sqrt{|A|^2+|B|^2}$. Since $N=R^2+2|C|^2$ is constant, each heteroclinic
orbit is a semi-ellipse. Motion starting on one of these semi-ellipses will
move towards an end-point, taking infinite time to reach it.

In Fig.~3 taken from Holm and Kova\v ci\v c 1992, we present another view of
the trajectories for $J=0$.  The Hamiltonian is
$$
H = \half (Z+Z^*) \cdot (\half N - |Z|^2)  \, .
$$
Accessible points lie on or within the circle $|Z|^2=N/2$. For $H=0$
the trajectory is the segment of the imaginary axis within the circle.
This is the  homoclinic orbit.
For $H\ne 0$, we solve for the imaginary part of $Z=Z_1+iZ_2$,
$$
Z_2 =  \pm \sqrt{ -Z_1^2+\half N - (H/Z_1) }
$$
This allows us to plot the trajectories for the range of $H$ for which
real solutions exist. There are two equilibrium points,
at $Z=\pm\sqrt{N/6}$, corresponding to solutions for which there is no
exchange of energy between the
vertical and horizontal components. These are the cup-like and cap-like
solutions first discussed by Vitt and Gorelik (1933).

\subsection{Geometry of the motion for fixed $J$}

The vertical amplitude is governed by equation (\ref{Qeqn}), which we write as
\begin{equation}
\frac{1}{2}\left(
\frac{ d{\cal Q} }{d\tau} \right)^2
        + {\cal V}({\cal Q}) = {\cal E}  \,,
\label{ENEQ}
\end{equation}
with the potential ${\cal V}({\cal Q})$ given by
\begin{equation}
{\cal V}({\cal Q}) =
    -\half\left[ {\cal Q}^3-2{\cal Q}^2+(1-{\cal J}^2){\cal Q}\right] \, .
\label{PotV}
\end{equation}
We note that ${\cal V}({\cal Q})$ has three zeros,
${\cal Q}=0$, ${\cal Q}=1-{\cal J}$ and ${\cal Q}=1+{\cal J}$.
Eqn. (\ref{ENEQ}) is an energy equation for a particle of
unit mass, with position ${\cal Q}$ and energy ${\cal E}$, moving in a cubic
potential field ${\cal V}({\cal Q})$.
In Fig.~4 we plot $\dot {\cal Q}$, given by (\ref{ENEQ}), against
${\cal Q}$ for the cases ${\cal J}=0$ (left panel) and ${\cal J}=0.25$
(right panel), for a range of values of ${\cal E}$.
Each curve represents the projection onto the reduced phase-space
of the trajectory of the modulation envelope.
The centers are relative equilibria, corresponding to the elliptic-parabolic
solutions of (Lynch, 2001), which are generalizations of the
cup-like and cap-like solutions of Vitt and Gorelik (1933).
The case ${\cal J}=0$ includes the homoclinic trajectory, for which $H=0$.

\subsection{Geometry of the motion for $H=0$}

For arbitrary ${\cal J}$, the $H=0$ motions are on a surface in the
space with  coordinates (${\cal Q}$, $\dot {\cal Q}$, $\cal J$).
This surface is depicted in Fig.~5. It has three singular points
(i.e., it is equivalent to a sphere with three pinches)
and its shape is similar to a tricorn hat.
The motion takes place on an intersections of this surface with a plane
of constant $\cal J$. There are three equilibrium solutions:
that with ${\cal J}=0$ (marked H.P.~in Fig.~5)
is at the extremity of the homoclinic
trajectory, and corresponds to purely vertical oscillatory
motion; those with ${\cal J}=\pm 1$ correspond to purely horizontal
motion, clockwise or anti-clockwise, with the spring tracing out a
cone. The purely vertical motion is unstable; the conical motions are
stable (perturbations about  conical motion were investigated by Lynch,
2001). The dynamics on the tricorn are similar to the motion of a
free rigid body. The three singular points  correspond to the
steady states of rotation about the three principal axes.

\subsection{Three-wave surfaces}

There is another way to depict the motion in reduced phase-space.
Let us consider a reduced phase-space with $x$ and $y$ axes
$X=\Re\{ABC^*\}$ and $Y=\Im\{ABC^*\}$ and $z$-axis ${\cal Q}=2|Z|^2/N$.
We note that $X \equiv H$.  It follows from
(\ref{Hconstant})--(\ref{Jconstant}) that
$$
X^2 + Y^2 = |A|^2|B|^2|C|^2 = \frak{1}{4}|Z|^2\Big[(2|Z|^2-N)^2 - J^2\Big] \, .
$$
We define ${\cal X} = (2/N^{3/2})X$ and ${\cal Y} = (2/N^{3/2})Y$ and can write
\begin{equation}
{\cal X}^2 + {\cal Y}^2
= \half\left[ {\cal Q}^3-2{\cal Q}^2+(1-{\cal J}^2){\cal Q}\right]
= -{\cal V}({\cal Q}) \, .
\label{TWsurf}
\end{equation}
where ${\cal V}$ is defined in (\ref{PotV}).
We note that ${\cal X}^2=-{\cal E}$ and
${\cal Y}^2=\half(d{\cal Q}/d\tau)^2$.
Eqn. (\ref{TWsurf}) implies that
the motion takes place on a surface of revolution about the
${\cal Q}$-axis. The radius for a given value of ${\cal Q}$ is the
square-root of the cubic $-{\cal V}({\cal Q})$.
The physically assessable region is $0 \le {\cal Q} \le 1-|{\cal J}|$.
Several such surfaces (for ${\cal J} \in \{0.0,0.1,0.2,0.3\}$) are shown
in Fig.~6. Since ${\cal X}^2={\cal H}^2=4H^2/N^3$, the motion for given
${\cal J}$ takes place on the intersection of the corresponding surface
of revolution with a plane of constant ${\cal X}$.

We can relate the tricorn surface to the surface of revolution.
The former is appropriate for $H=0$; the $H\ne 0$ case
is represented by trajectories inside this surface.
If we slice the tricorn surface in a plane of fixed ${\cal J}$
we get a set of closed trajectories, the outside one
for $H=0$ the others for $H\ne 0$
(the cases ${\cal J}=0$ and ${\cal J}=0.25$ are plotted in Fig.~4 above.)
If we now distort the ${\cal J}$-section into a cup-like surface, by taking
$H$ as a vertical coordinate and plotting each trajectory at
a height depending on its $H$ value, we get half of a closed surface.
Each trajectory is selected by a $H$-plane section.  Alteration of the sign
of $H$ corresponds to reversal of time.  Completing
the surface by reflection in the plane $H=0$ gives the surface
generated by rotating the root-cubic graph
$\sqrt{-{\cal V}({\cal Q})}$ about the ${\cal Q}$-axis,
{\it i.e.}, the surface given by (\ref{TWsurf}).
These surfaces are what Alber, {\it et al.}, (1998a)
call the three-wave surfaces. They foliate the volume contained within the
surface for ${\cal J}=0$.


\section{The Precession of the swing plane}

The characteristic feature of the behavior of the physical
spring is its stepwise precession, which we shall now analyze.
As the oscillations change from horizontal to vertical and back again,
it is observed that each successive horizontal excursion departs in a
different direction. The only reference to this phenomenon of which we are
aware,  prior to Lynch (2001), is Cayton (1977). Cayton briefly discussed
three-dimensional solutions and mentioned the precession of the swing
plane, but did not analyze its dynamics. Surprisingly, the characteristic
stepwise precession of the swinging spring has been largely ignored,
although it
is immediately obvious upon observation of a physical elastic pendulum  with
$\omega_Z \approx 2\omega_R$. Indeed, this precession is almost impossible to
suppress experimentally when the initial motion is close to vertical.

\subsection{Qualitative description}

If the horizontal projection of the motion is an ellipse of high
eccentricity, the motion is approximately planar.  We call the vertical plane
through the major axis of this ellipse the {\it swing plane}.  When the
initial
oscillations are quasi-vertical, the motion gradually develops into an
essentially horizontal swinging motion. This horizontal swinging does
not persist, but soon passes again into nearly vertical springing
oscillations similar to the initial motion. Subsequently, a horizontal swing
again develops, but now in a different direction.  The stepwise precession of
this exchange between springing and swinging motion continues indefinitely in
the absence of dissipation and is the characteristic experimental feature
of the
swinging spring. We shall seek an expression for the change in direction of the
swing plane from one horizontal excursion to the next.

A full knowledge of the solutions of the three equations of
motion would of course suffice to determine the swing plane at each moment in
time. In Lynch (2001) the equations were expressed in rotating
co-ordinates, and a particular solution for the slow rotation of the
swing plane was posited as a function of the vertical amplitude $|C|$,
by assuming a certain angular relation. Following this assumption, the
angle of the swing-plane could be expressed as an integral involving
elliptic functions.

\subsection{Pattern evocation in shape space}

We shall approach the precession problem using pattern evocation in
shape space. Pattern evocation seeks a relative equilibrium
(in shape space) in which a phase relationship between the variables
(the shape) is preserved (Marsden, {\it et al.}, 1995, 1996).
We track the pattern by moving to a non-uniformly
rotating frame in which the orientation of the shape is fixed.
This is a generalization of the idea of tracking a satellite orbit by
evoking constancy of the areal velocity required to conserve angular
momentum.

Our particular geometric assumption is that the angle between the complex
amplitudes $a$ and $b$ remains constant, in an appropriately rotating frame.
Writing these amplitudes in vector form as ${\bf a} =
(|a|\cos\alpha,|a|\sin\alpha,0)$,
${\bf b} = (|b|\cos\beta,|b|\sin\beta,0)$ and taking ${\bf k} = (0,0,1)$ yields
$$
J = -\,{\bf k \cdot a \times b} = |ab|\sin(\alpha-\beta) \, , \qquad
{\bf a \cdot \bf b} = |ab|\cos(\alpha-\beta)
\, .
$$
Consequently, our geometric pattern evocation assumption that the phase
difference $\alpha-\beta$ remains constant immediately implies that $|ab|$ is
also constant. The conservation of angular momentum $J$ means that the area of
the  parallelogram formed by the vectors ${\bf a}$ and ${\bf b}$ is constant.
The requirement of constant $\alpha-\beta$ imposes an additional
geometric constraint on the possible shape of the orbits. For
example, when $\alpha-\beta=\pi/2$ (mod$\pi$), the orbits are elliptical.

\subsection{Modulation equations in rotating coordinates}

We shall transform to rotating co-ordinates and seek an expression for the
(slow) rotation frequency $\Omega(t)$ that allows us to estimate the stepwise
precession of the swinging spring by imposing the pattern evocation constraint
that $\alpha-\beta$ remains constant.

In a {\bf rotating frame}, the approximate Lagrangian (\ref{Lag-3D-0}) at cubic
order in the coordinate displacements becomes, with $\mathbf{x}=(x,y,z)$,
\begin{equation}\label{Lag-rot}
L = \half |\dot{\mathbf{x}} + \Omega(t)\,\hat{\mathbf{z}}\times\mathbf{x}|^2
 -  \half \left( \omega_R^2(x^2+y^2)+\omega_Z^2 z^2 \right)
 +  \half \lambda(x^2+y^2) z
\,.
\end{equation}
Now $x$, $y$ and $z$ are Cartesian coordinates centered at the point
of equilibrium {\it in the rotating frame}, $\omega_R=\sqrt{g/\ell}$ is the
frequency of linear pendular motion, $\omega_Z=\sqrt{k/m}$ is the frequency of
its elastic oscillations and $\lambda=\ell_0\omega_Z^2/\ell^2$.
The corresponding Euler-Lagrange equations of motion
(\ref{XXeqn})--(\ref{ZZeqn}) may be written in rotating coordinates as
\begin{eqnarray}
\ddot x - \dot{\Omega}(t)y - 2\Omega(t)\dot{y}
+ \big(\omega_R^2 - \Omega^2(t)\big) x
&=&
\lambda  xz
\,,    \label{XXeqn-rot}
\\
\ddot y + \dot{\Omega}(t)x + 2\Omega(t)\dot{x}
+ \big(\omega_R^2 - \Omega^2(t)\big) y
&=&
\lambda yz
\,,    \label{YYeqn-rot}
\\
\ddot z + \omega_Z^2 z
&=&
\half\lambda (x^2+y^2)
\, .
\label{ZZeqn-rot}
\end{eqnarray}
The vertical component of angular momentum is
$$
h = \hat{\mathbf{z}}\cdot \mathbf{x} \times
\Big(\dot{\mathbf{x}} + \Omega(t)\,\hat{\mathbf{z}}\times\mathbf{x}\Big)
  = (x\dot y - \dot x y) + \Omega(t)(x^2+y^2)
$$
and is a constant of the motion for these equations.
However, upon Legendre-transforming, one finds the time-dependent Hamiltonian
satisfies
$$
\dot{H} = - \, \dot{\Omega}(t) h
\,.
$$
Thus, perhaps not unexpectedly, exact energy conservation breaks down to the
extent that the rotation frequency is nonuniform.

\subsection{Averaged Lagrangian and modulation equations for slow rotation}

The modulation equations in axes rotating with angular velocity $\Omega(t)$
about the vertical are obtained in the resonant case $\omega_Z=2\omega_R$ by
applying the averaged Lagrangian technique (Whitham, 1974). Accordingly, the
solution of (\ref{XXeqn-rot})--(\ref{ZZeqn-rot}) is assumed to be of the form
\begin{eqnarray}
x &=& \Re[a(t)\exp(i\omega_R t)] \, ,  \label{ModSolA-R} \\
y &=& \Re[b(t)\exp(i\omega_R t)] \, ,  \label{ModSolB-R} \\
z &=& \Re[c(t)\exp(2i\omega_R t)] \,   \label{ModSolC-R}
\,.\end{eqnarray}
(Note that subscript-zeroes are dropped for these modulation amplitudes in
the rotating frame.) In these variables, the averaged Lagrangian corresponding
to (\ref{Lag-rot}) may be written as
\begin{eqnarray}\label{Lag-rot-mod}
\langle L \rangle
&=&
\half\omega_R
\Big[  \Im\{\dot a a^* + \dot b b^* + 2\dot c c^*\}
 + \Re\{(a^2+b^2)c^*\} + 2\Omega \Im\{ab^*\} \Big]
\nonumber\\
&&\hspace{20mm}
+\
\half\Omega\,\Re\,\big[a^*\dot{b} - \dot{a}^*b\big]
+
\frak{1}{4}\Omega^2
\Big[|a|^2+|b|^2\Big]
\,.
\end{eqnarray}
On
assuming that the rotation frequency is sufficiently slow that
$\Omega/\omega_R\ll1$, we shall {\it neglect} all terms in the averaged
Lagrangian (\ref{Lag-rot-mod}) that are not multiplied by $\omega_R$. In this
approximation of slow rotation, the averaged Lagrangian is given by
the simpler expression,
\begin{equation}\label{Lag-rot-mod-approx}
\langle L \rangle
=
\half\omega_R
\Big[  \Im\{\dot a a^* + \dot b b^* + 2\dot c c^*\}
 + \Re\{(a^2+b^2)c^*\} + 2\Omega J \Big]
\,.
\end{equation}
Here $J=\Im\{ab^*\}$ is the angular momentum,
a conserved quantity at this level of approximation and
formally identical to the expression in non-rotating coordinates.
The Euler-Lagrange modulation equations in this approximation may be
written as
\begin{eqnarray}
i\dot a = a^*c + i\Omega b
\,,\label{Adot-rot} \\
i\dot b = b^*c - i\Omega a
\,,\label{Bdot-rot}\\
i\dot c = \frak{1}{4}(a^2+b^2)
\label{Cdot-rot}
\,.\end{eqnarray}
We may also write these leading order
equations in Hamiltonian form. When
$\langle H \rangle$ is defined by
$$
\langle H \rangle = \Re\{(a^2+b^2)c^*\} + 2\Omega\Im\{ab^*\}
\,,
$$
with coordinates ($a,b,c$), conjugate momenta ($a^*,b^*,2c^*$) and
Poisson brackets defined by $\{a,a^*\}=\{b,b^*\}=2\{c,c^*\}=-i$, the modulation
equations (\ref{Adot-rot})--(\ref{Cdot-rot}) are expressible in canonical
Hamiltonian form as,
$$
i\dot a = i\{a,\langle H \rangle\}
        = \frac{\partial \langle H \rangle }{ \partial {a^*}} \, , \quad
i\dot b = i\{b,\langle H \rangle\}
        = \frac{\partial \langle H \rangle }{ \partial {b^*}} \, , \quad
i\dot c = i\{c,\langle H \rangle\}
        = \frac{\partial \langle H \rangle }{ \partial {\,2c^*}} \, .
$$
The three constants of the motion for these equations are
\begin{eqnarray}
H_0 &=& \half[(a^2+b^2)c^*+({a^*}^2+{b^*}^2)c]
=
\Re \{(a^2+b^2)c^*\}
\,,\\
N   &=& |a|^2 + |b|^2 + 4|c|^2
\,,\\
J   &=& (ab^*-a^*b)/2i
=
\Im\{ab^*\}
\,.\end{eqnarray}

We now introduce the pattern evocation assumption.  Noting that
$$
|ab|^2 = (\Re\{ab^*\})^2 + (\Im\{ab^*\})^2 \, ,
$$
and that the second term is just $J^2$, implies
constancy of $\Re\{ab^*\}$.
Using Eqns. (\ref{Adot-rot}) and (\ref{Bdot-rot}) it follows that
\begin{equation}
\frac{ d}{dt}|ab|^2 =
-2\Re\{ab^*\}\left[  2\Im\{abc^*\} + \Omega(|a|^2-|b|^2)  \right] = 0 \, .
\label{Factors}
\end{equation}
Either factor may vanish, so there appear to be two possibilities
for the solution.  We first assume that the factor in square brackets
in (\ref{Factors}) vanishes. This implies
\begin{equation}
\Omega =  -\,\frac{ 2\Im\{abc^*\} }{ |a|^2-|b|^2 }
       =  -\,\frac{ 2|abc|\sin(\alpha+\beta-\gamma) }{ |a|^2-|b|^2 } \, .
\label{OmegaGen}
\end{equation}
(where $c=|c|e^{i\gamma}$).
The precession angle $\Theta=\int_0^t\Omega(t')dt'$ can be ascertained by
integrating
$\Omega$ over the time interval of the motion.
In the special case $\alpha-\beta=\frak{\pi}{2}$(mod$\,\pi$), one finds by
using
the constants of motion that
\begin{equation}
\Omega =  \frac{ 2J H_0  }{ (N-4|c|^2)^2-4J^2 } \, .
\label{OmegaSpec}
\end{equation}
This case also corresponds to the vanishing of the first factor in
(\ref{Factors}), so that $\Re\{ab^*\}=0$ and
$\bf a$ and $\bf b$ are $90^{\textstyle{\circ}}$ out of phase.
This was the Ansatz introduced by Lynch (2001).
He showed that, in this case, the rotation rate is given by (\ref{OmegaSpec}).
We now see that the result in Lynch (2001) is a special case of
the general result (\ref{OmegaGen}).
In this special case, $\Omega$ can be computed as soon as $|c|$ is known.
We will examine this case numerically below.


\subsection{The instantaneous ellipse}

In order to define precisely the precession angle,
we introduce an ellipse which approximates the
horizontal projection of the trajectory of the pendulum.
Recall that the full solution for the horizontal components is
$$
x = \Re\{a\exp(i\omega_R t)\} = |a|\cos(\omega_R t+\alpha) \, , \qquad
y = \Re\{b\exp(i\omega_R t)\} = |b|\cos(\omega_R t+\beta ) \, ,
$$
where $\alpha$ and $\beta$ are the phases of $a$ and $b$.
The amplitudes and phases are assumed to vary slowly.
If they are regarded as constant over a period $\tau=1/\omega_R$ of the
fast motion, these equations describe a central ellipse,
\begin{equation}
Px^2 + 2Qxy + Ry^2 = S    \label{InstEll}
\end{equation}
where $P=|b|^2$, $Q=-|ab|\cos(\alpha-\beta)$, $R=|a|^2$ and $S=J^2$.
The area of the ellipse is easily calculated and is found to have the
constant value $\pi J$.
Its orientation is determined by eliminating the cross-term in
(\ref{InstEll}). This is achieved as usual by rotating the axes
through an angle $\theta$, given by
\begin{equation}
\tan 2\theta = \frac{ 2Q }{ P-R } =
               \frac{ 2|ab|\cos(\alpha-\beta) }{ |a|^2-|b|^2 } \, .
\label{ElAzimuth}
\end{equation}
The semi-axes of the ellipse are given by
\begin{equation}
A_1 = \frac{ J }{ \sqrt{ P\cos^2\theta + Q\sin 2\theta + R\sin^2\theta } } \, ,
\quad
A_2 = \frac{ J }{ \sqrt{ P\sin^2\theta - Q\sin 2\theta + R\cos^2\theta } }
\, .
\label{SemiAxes}
\end{equation}
The area is $\pi A_1A_2 = \pi J$ and the eccentricity can be calculated
immediately. In the case of unmodulated motion, such as the elliptic-parabolic
modes  described in Lynch (2001), the instantaneous ellipse corresponds to the
trajectory, which is a precessing ellipse.
In general, it is only an approximation to the trajectory, but we may define
the orientation or azimuth at any time to be the angle $\theta$ given by
(\ref{ElAzimuth}). This angle will be compared to the precession angle $\Theta$
calculated by integrating (\ref{OmegaSpec}) and shown to give almost identical
results.



\def\Degree{^{\textstyle{\circ}}}

\section{Numerical results}

We examine the results of numerical integrations of the modulation equations
(\ref{niceone})--(\ref{nicethree}) and compare them to the solutions of the
exact equations (\ref{XXeqn})--(\ref{ZZeqn}).
It will be seen that the modulation equations provide an excellent
description of the envelope of the rapidly varying solution of the full
equations.
We then compare the stepwise precession angle predicted by a formula based
on constancy of the angle $\alpha-\beta$ with the numerical
simulation of this quantity and show that the two values track each
other essentially exactly.

The parameter values chosen for all numerical integrations are
$m=1\,$kg,\ $\ell=1\,$m,\ $g=\pi^2\,\rm m\,s^{-2}$ and $k=4\pi^2\,\rm
kg\,s^{-2}$ so that $\omega_R=\pi$, $\omega_Z=2\pi$ and the resonance
condition $\omega_Z=2\omega_R$ holds.
The linear rotational mode has period $\tau_R=2\,$s, and the vertical
mode has period $\tau_Z=1\,$s. The initial conditions are set as follows,
$$
(x_0, y_0, z_0) = ( 0.006, 0, 0.012 )\, ; \qquad
(\dot x_0, \dot y_0, \dot z_0) = (0, 0.00489, 0) \, .
$$
(The value of $\dot y_0$ was chosen to tune the precession angle
to be an even fraction of $180\Degree$, making the amplitudes,
though not the phases, periodic).
The corresponding initial values for the modulation
equations (\ref{niceone})--(\ref{nicethree}) are given by
$$
\alpha_0 = \arctan\left(\frac{-\dot x_0}{  \omega_R x_0}\right)
\, , \quad
\beta_0  = \arctan\left(\frac{-\dot y_0}{  \omega_R y_0}\right)
\, , \quad
\gamma_0 = \arctan\left(\frac{-\dot z_0}{ 2\omega_R z_0}\right)
\, ,
$$
$$
|a_0| =  \left(\frac{x_0}{\cos\alpha_0 }\right)\,, \quad
|b_0| = -\left(\frac{\dot y_0}{\omega_R\sin\beta_0}\right) \,, \quad
|c_0| =  \left(\frac{z_0}{\cos\gamma_0}\right) \,,
$$
giving the values $(|a_0|, |b_0|, |c_0|) = (0.006,0.002,0.012)$ and
$(\alpha_0,\beta_0,\gamma_0)= (0,-\pi/2,0)$.
The constants of the motion take the following values,
$$
H = 4.03\times 10^{-7}
\,,\qquad
J = 9.34\times 10^{-6}
\,,\qquad
N = 6.14\times 10^{-4}
\,.
$$
The integration was extended over a period of 1000 seconds (i.e., 1000
vertical oscillations). As a check on numerical accuracy, the changes in these
quantities, which should remain constant, was calculated, with the following
results:
$$
\left( \frac{ H_{\rm Final} }{ H_{\rm Initial} } \right) = 100.04\%  \,, \quad
\left( \frac{ J_{\rm Final} }{ J_{\rm Initial} } \right) = 99.997\%  \,, \quad
\left( \frac{ N_{\rm Final} }{ N_{\rm Initial} } \right) = 100.00\%   \,.
$$

We now directly compare the solutions of the `exact' equations
(\ref{XXeqn})--(\ref{ZZeqn}) and the `approximate' or modulation
equations (\ref{niceone})--(\ref{nicethree}).
Once the modulation equations have been solved for the envelope
amplitudes and phases, the full approximate solution is given by
(\ref{ModSolA})--(\ref{ModSolC}).
We first consider the horizontal projection of the solution for the
1000 second integration. This is the period required for the solution
to precess through approximately $180\Degree$.
In Fig.~7 (top panel) we plot $x$ versus $y$ for the exact solution.
In Fig.~7 (bottom panel) we plot the corresponding solution from the modulation
equations.  It is clear that there is great similarity between the two
solutions;
indeed, the two plots are indistinguishable.
The precession angle between horizontal excursions is close to $30\Degree$
(the value of $\dot y_0$ was chosen to ensure this). The modulation period
is approximately 167 seconds; thus, the instantaneous ellipse rotates
through six cycles and $180\Degree$ in 1000 seconds.

The vertical structure of the solution is displayed in Fig.~8, where
$z$ for the exact solution (top panel) and $\Re\{c_0(t)\exp(2i\omega_R t)\}$
for the approximate solution (bottom panel) are seen to be virtually identical.
For clarity, the solutions are plotted only for the first modulation cycle
of 167 seconds.  The character of the solution --- rapid oscillations with a
slowly-varying amplitude envelope --- is clear from the figure. The vertical
amplitude is close to zero when horizontal excursions are at their peak.
This is confirmed in Fig.~9 (top panel) where the horizontal modulation
amplitude
$S = \sqrt{|a|^2+|b|^2}$ and vertical modulation amplitude $C=|c|$ are plotted
against time.

In Fig.~9 (bottom panel)
we plot the squared eccentricity $e^2 = (1-A_{\rm min}^2/A_{\rm maj}^2)$
of the envelope of the horizontal projection of the
approximate solution, where the semi-axes $A_{\rm maj}$ and
$A_{\rm min}$ are calculated from (\ref{SemiAxes}).
The eccentricity is close to
unity for most of the integration.
Horizontal excursions of the pendulum occur during this time.
For short periods, when the horizontal amplitude is minimum, the value of $e$
drops significantly (solid line). During this time, the angular velocity,
calculated as the rate of change of the azimuth given by (\ref{ElAzimuth}),
reaches a maximum (dashed line).
Thus, the precession occurs in bursts near the times when the vertical
amplitude
is maximum and horizontal amplitude minimum.

The stepwise nature of the precession is clearly illustrated in Fig.~10.
The azimuthal angle $\vartheta$ of the numerical solution of the exact
equations may be calculated by fitting a central conic to every three
consecutive points on the trajectory. Assuming a solution of the form
$$
\tilde Px^2 + 2\tilde Qxy + \tilde Ry^2 = 1    \label{ExEllip}
$$
and requiring that the three points lie on this curve, we obtain three
equations for the coefficients  $(\tilde P, \tilde Q, \tilde R)$.
>From these, the azimuth $\vartheta$ and the semi-axes are obtained from
equations analogous to (\ref{ElAzimuth}) and (\ref{SemiAxes}).
This is compared in Fig.~10 to the corresponding value $\theta$ resulting from
integration of the modulation equations.
It is noteworthy that $\vartheta$ and $\theta$ remain quasi-constant for most
of the modulation cycle, changing rapidly only over  short intervals around the
times when $C$ is maximum and $S$ is minimum. The advances in phase are very
similar for the exact and approximate solutions. However, there are small
differences: $\theta-\vartheta$ is also plotted in Fig.~10 (dotted line). This
sensitive quantity reaches its maximum value of
$4.35\Degree$ at the end of the integration.

The azimuthal change between successive horizontal excursions is very close to
$30\Degree$ for both exact and approximate solutions.
We also calculated the angle $\Theta$ resulting from an
integration of (\ref{OmegaSpec}). The graphs of $\theta$ and $\Theta$
(not plotted) are indistinguishable. The maximum difference $|\theta-\Theta|$
was only $0.0063\Degree$. This is remarkable: the value $\Theta$ derived from
(\ref{OmegaSpec}) involves an assumption that $\alpha-\beta$ is constant
in a particular rotating frame.
The azimuth $\theta$ from the modulation equations makes no such assumption,
yet the two solutions are practically identical. This confirms that the
pattern evocation assumption which yields the result (\ref{OmegaSpec}) is
sound.

Numerous other integrations of the exact and modulation equations were
also carried out. They confirm that the stepwise precession of the azimuthal
angle is a distinct characteristic of the swinging spring. This is also
in complete agreement with simple experiments with a physical pendulum,
where the periodic exchange of energy between horizontal and vertical
and the precession of the swing plane between horizontal excursions are
the main observable properties of the motion.


\section*{Acknowledgements}

The authors would like to acknowledge the facilities provided by the
Isaac Newton Institute at Cambridge University, where this work began. Peter
Lynch would like to thank the  Center for Nonlinear Studies, Los Alamos
National Laboratory, for support and hospitality. Darryl Holm is grateful to
Gregor Kova\v ci\v c and Jerry Marsden for many fruitful discussions and
ongoing collaborations in dynamical systems analysis. DDH was supported by US
DOE, under contract W-7405-ENG-36.



\section*{References}

{\small{

\parskip 10pt
\parskip  0pt

\frenchspacing

\refce
Aceves, A B, D D Holm, G Kova\v ci\v c and I Timofeyev, 1997: Homoclinic
orbits and chaos in a second-harmonic generating optical cavity.
{\it Phys. Lett. A} {\bf233}, 203--208.

\refce
Alber, M S, G G Luther, J E Marsden and J M Robbins, 1998a:
Geometric phases, reduction and  Lie-Poisson structure for the resonant
three-wave interaction,
{\it Physica D} {\bf 123}, 271--290.

\refce
Alber, M S, G G Luther, J E Marsden and J M Robbins, 1998b:
Geometry and Control of Three-Wave Interactions.
({\it Caltech preprint}\/)

\refce
Bretherton, F P, 1964:
Resonant interactions between waves. The case of discrete oscillations.
{\it J~Fluid~Mech.}, {\bf 20}, 457-479.

\refce
Cayton, Th~E, 
The laboratory spring-mass oscillator: an example of parametric
instability.
{\it Am.~J.~Phys.}, {\bf 45}, 723--732 (1977).

\refce
David, D and D D Holm:
Multiple Lie-Poisson Structures, Reductions, and
Geometric Phases for the Maxwell-Bloch Traveling-Wave Equations,
{\it J. Nonlin. Sci.} {\bf 2} (1992)  241--262.

\refce
David, D, D D Holm and M Tratnick:
Hamiltonian Chaos in Nonlinear
Optical Polarization Dynamics,
{\it Phys. Rep.} {\bf 187} (1990) 281--367.

\refce
Hasegawa, A and K~Mima, 1977:
Pseudo-three-dimensional turbulence in magnetized nonuniform plasmas.
{\it Phys.~Fluids}, {\bf 21}(1), 87--92.

\refce
Holm, D D and G Kova\v ci\v c, 1992:
Homoclinic chaos in a laser-matter system.
{\it Physica D}  {\bf 56}, 270--300.

\refce
Holm, D D, G Kova\v ci\v c and T A Wettergren, 1995:
Near integrability and chaos in a resonant-cavity laser model.
{\it Phys. Lett. A}, {\bf 200}, 299--307.

\refce
Holm, D D, G Kova\v ci\v c and T A Wettergren, 1996:
Homoclinic orbits in the Maxwell-Bloch equations with a probe.
{\it Phys. Rev. E}, {\bf 54}, 243--256.

\refce
Holm, D D and J E Marsden, 1991:
The rotor and the pendulum.
in {\it Symplectic Geometry and  Mathematical Physics},
P. Donato, C. Duval, J. Elhadad, G. M. Tuynman, ed.,
Prog. in Math. Vol. {\bf 99},
Birkhauser: Boston, 1991, pp. 189--203.

\refce
Horton, W and A~Hasegawa, 1994:
Quasi-two-dimensional dynamics of plasmas and fluids.
{\it Chaos}, {\bf 4}(2), 227--251.

\refce
Longuet-Higgins, M S and A E Gill, 1967:
Resonant interactions between planetary waves
{\it Proc.~Roy.~Soc.}, {\bf A299}, 120--140.

\refce
Lorenz, E N, 1963:
Deterministic non-periodic flow.
{\it J.~Atmos.~Sci.}, {\bf 20}, 130--141.

\refce
Luther, G~G, M~S~Alber, J~E~Marsden and J~M~Robbins, 2000:
Geometric analysis of optical frequency conversion and its control in
quadratic nonlinear media.
{\it J.~Optical~Soc.~Am.}, {\bf B/17}, 932-941.

\refce
Lynch, P, 2001:
Resonant motions of the three-dimensional elastic pendulum.
To appear in {\it Intl.~J.~Nonlin.~Mech.} (2001).

\refce
Marsden, J E, J Scheurle and J M Wendlandt, 1996:
Visualization of orbits and pattern evocation for the
double spherical pendulum.
ICIAM 95 (Hamburg, 1995), 213--232,
{\it Math. Res.}, {\bf 87}, Akademie Verlag, Berlin, 1996.

\refce
Marsden, J E and J Scheurle, 1995:
Pattern evocation and geometric phases in mechanical
systems with symmetry.
{\it Dynam.~Stability Systems}, {\bf 10}(4), 315--338.

\refce
Montgomery, R, 1991:
How much does the rigid body rotate? A Berry's phase from the 18th
century.
{\it Am.~J.~Phys.}, {\bf 59}(5), 394--398.

\refce
Ott, Edward, 1993:
{\it Chaos in Dynamical Systems}.
Cambridge Univ.~Press, 385pp.

\refce
Pedlosky, J, 1987:
{\it Geophysical Fluid Dynamics}.
Springer-Verlag, 710pp.

\refce
Sparrow, C., 1982:
{\it The Lorenz Equations: Bifurcations, Chaos and Strange Attractors.}
Springer-Verlag, New York, 269pp.

\refce
Vitt, A and G~Gorelik, 
Kolebaniya uprugogo mayatnika kak primer kolebaniy
dvukh parametricheski svyazannykh linejnykh sistem.
{\it Zh.~Tekh.~Fiz.} ({\it J.~Tech.~Phys.}) {\bf 3}(2-3), 294--307
(1933).
Available in English translation:
{\it Oscillations of an Elastic Pendulum as
an Example of the Oscillations of Two
Parametrically Coupled Linear Systems}.
Translated by Lisa Shields, with an Introduction by Peter Lynch.
Historical Note No.~3, Met \'Eireann, Dublin (1999).

\refce
Wersinger, J-M, J~M~Finn and E~Ott, 1980:
Bifurcations and strange behavior in instability saturation
by nonlinear mode coupling.
{\it Phys.~Rev.~Lett.}, {\bf 44}, 453--456.

\refce
Whitham, G~N, 1974:
{\it Linear and Nonlinear Waves}, John Wiley \& Sons.

\nonfrenchspacing

}
}


\vfill\eject
\section*{Figure Captions}

\def\figcap{\smallskip \noindent\hangindent\parindent}

\figcap
Figure 1. Schematic diagram of the elastic pendulum or swinging spring.
Cartesian coordinates centered at the position of equilibrium are used.

\figcap
Figure 2. ${\cal C}$ is the plane of critical points, $A=0=B$.
The vertical axis is $R=\sqrt{|A|^2+|B|^2}$.
The vertical plane contains heteroclinic semi-ellipses
passing from $c_0$ to $-c_0$.

\figcap
Figure 3. Phase portrait in the $Z$-plane for $J=0$.
The motion is confined within the circle $|Z|^2=\half N$.
The segment of the imaginary axis within this circle is
the homoclinic orbit.

\figcap
Figure 4.  Plots of  $\dot {\cal Q}$ versus ${\cal Q}$
for ${\cal J}=0$ and ${\cal J}=0.25$, for a range of values
\hfill\break\qquad
$E \in \{ -0.0635, -0.0529, -0.0423, -0.0317, -0.0212, -0.0106, 0 \}$.

\figcap
Figure 5.  Tricorn surface, upon which motion takes place when $H=0$.
The coordinates are ${\cal J}$, ${\cal Q}$, $\dot {\cal Q}$.
The motion takes place on the intersections of this surface with a plane
of constant $\cal J$ (such planes are indicated by the stripes).
This surface has three singular points.
The homoclinic point is marked H.P.

\figcap
Figure 6. Surfaces of revolution about the ${\cal Q}$-axis, for
${\cal J}\in\{0.0,0.1,0.2,0.3\}$.
The radius for given ${\cal Q}$ is given
by the square-root of the cubic $-{\cal V}({\cal Q})$.
For given ${\cal J}$, the motion takes place on the intersection
of the corresponding surface with a plane of constant $X$.

\figcap
Figure 7. Horizontal projection of the solution for an integration
of 1000 seconds. Top: solution of the `exact' equations.
Bottom: solution of the `approximate' equations.

\figcap
Figure 8. Vertical amplitude of the solution for the
first modulation cycle (first 167 seconds).
Top: solution of the `exact' equations.
Bottom: solution of the `approximate' equations.

\figcap
Figure 9.  Top panel: Envelope amplitude of the approximate solution.
$S=\sqrt{|a|^2+|b|^2}$ (solid line) and $C=|c|$ (dashed line).
Bottom panel: Square of the eccentricity (solid line) and
angular velocity $\Omega$ (scaled by 50) of the instantaneous ellipse
(dashed line).

\figcap
Figure 10. Azimuth angle (in degrees) for the `exact' solution
($\vartheta$, solid line) and the `approximate' solution
($\theta$, dashed line).
The difference $\theta-\vartheta$ is plotted as a dotted line.
The azimuth $\Theta$ resulting from integration of (\ref{OmegaSpec})
(not plotted) is indistinguishable from the values $\theta$ of the
approximate solution.

\vfill\eject



%
%
%
%
%
%
\newdimen\rotdimen
\def\vspec#1{\special{ps:#1}}
\def\rotstart#1{\vspec{gsave currentpoint currentpoint translate
   #1 neg exch neg exch translate}}
\def\rotfinish{\vspec{currentpoint grestore moveto}}
%
%
\def\rotr#1{\rotdimen=\ht#1\advance\rotdimen by\dp#1%
   \hbox to\rotdimen{\hskip\ht#1\vbox to\wd#1{\rotstart{90 rotate}%
   \box#1\vss}\hss}\rotfinish}
%
%
\def\rotl#1{\rotdimen=\ht#1\advance\rotdimen by\dp#1%
   \hbox to\rotdimen{\vbox to\wd#1{\vskip\wd#1\rotstart{270 rotate}%
   \box#1\vss}\hss}\rotfinish}%
%
%
\def\rotu#1{\rotdimen=\ht#1\advance\rotdimen by\dp#1%
   \hbox to\wd#1{\hskip\wd#1\vbox to\rotdimen{\vskip\rotdimen
   \rotstart{-1 dup scale}\box#1\vss}\hss}\rotfinish}%
%
%
\def\rotf#1{\hbox to\wd#1{\hskip\wd#1\rotstart{-1 1 scale}%
   \box#1\hss}\rotfinish}%

%
%

\input epsf
\epsfverbosetrue

\newcount\Figsinline \def\Figsinline{0}

\ifnum\Figsinline=0
\hrule height 1pt
    \vfill
\includegraphics[scale=0.5,angle=0]{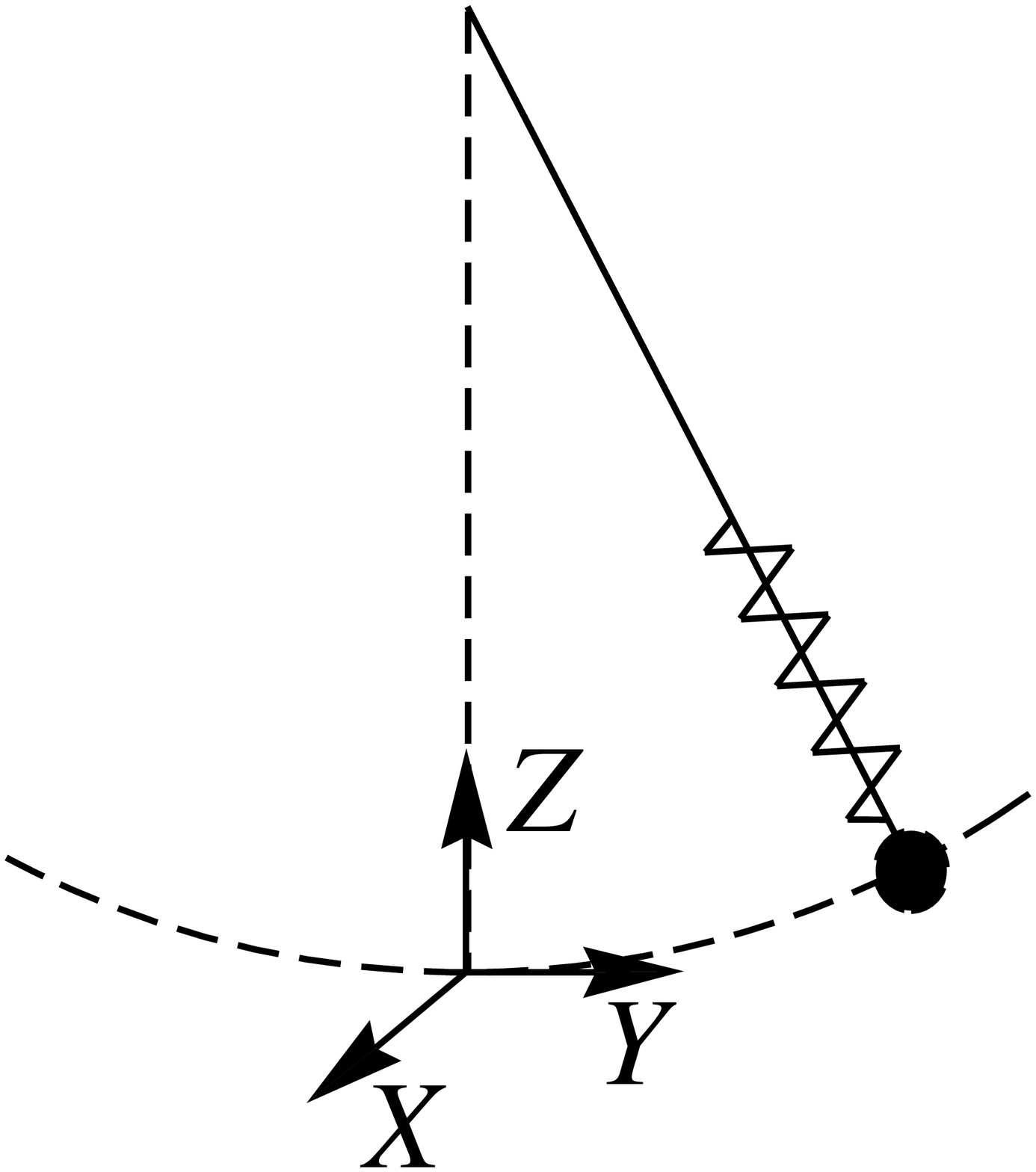}
      \vfill
\noindent
Figure 1. Schematic diagram of the elastic pendulum, or swinging spring.
Cartesian coordinates centered at the position of equilibrium are used.
      \vfill
\hrule height 1pt
\eject
\fi

\ifnum\Figsinline=0
\hrule height 1pt
    \vfill
\includegraphics[scale=0.6,angle=0]{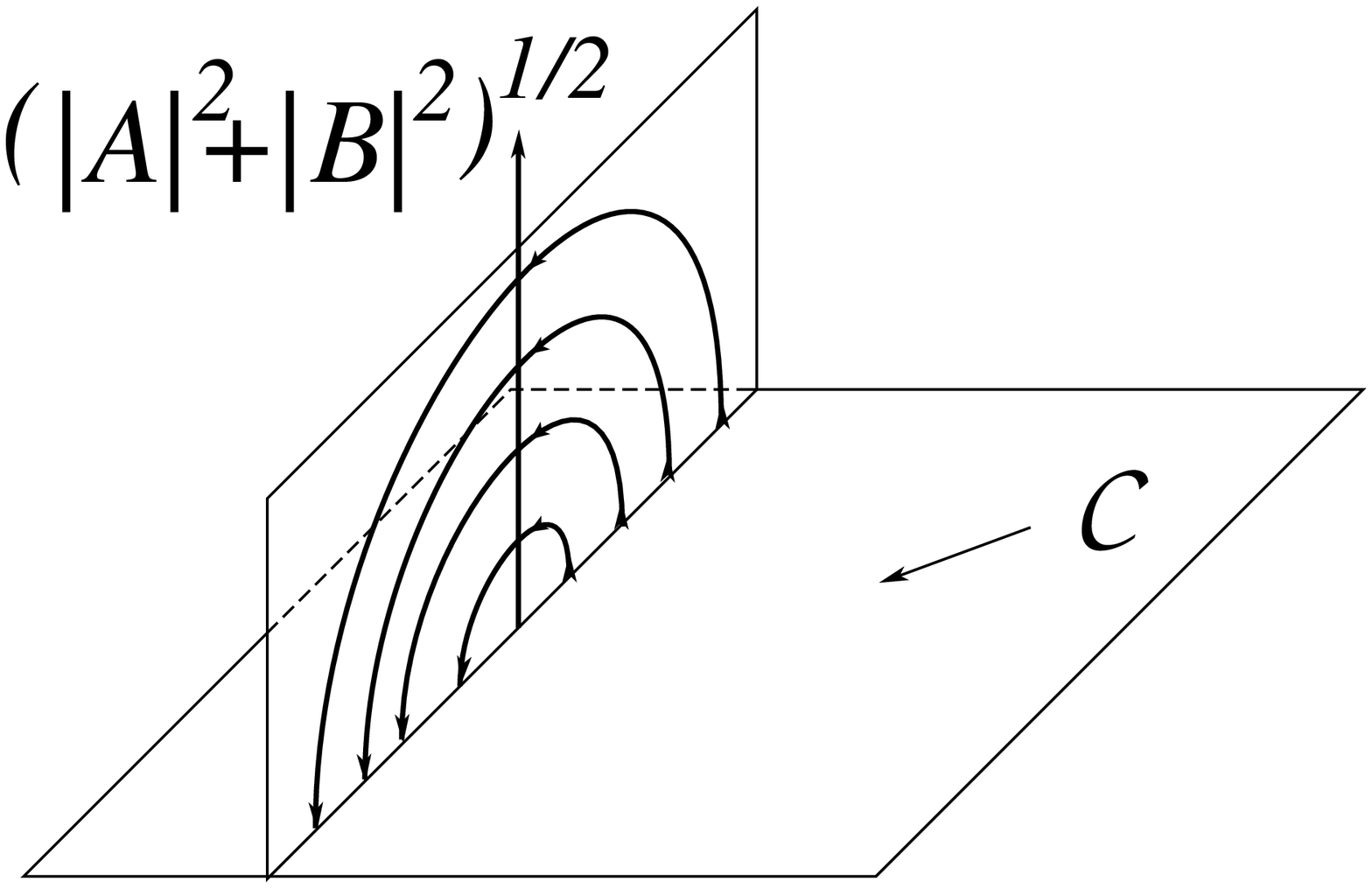}
      \vfill
\noindent
Figure 2. ${\cal C}$ is the plane of critical points, $A=0=B$.
The vertical axis is $R=\sqrt{|A|^2+|B|^2}$.
The vertical plane contains heteroclinic semi-ellipses
passing from $c_0$ to $-c_0$.
      \vfill
\hrule height 1pt
\eject
\fi

\ifnum\Figsinline=0
\hrule height 1pt
    \vfill
\includegraphics[scale=0.6,angle=0]{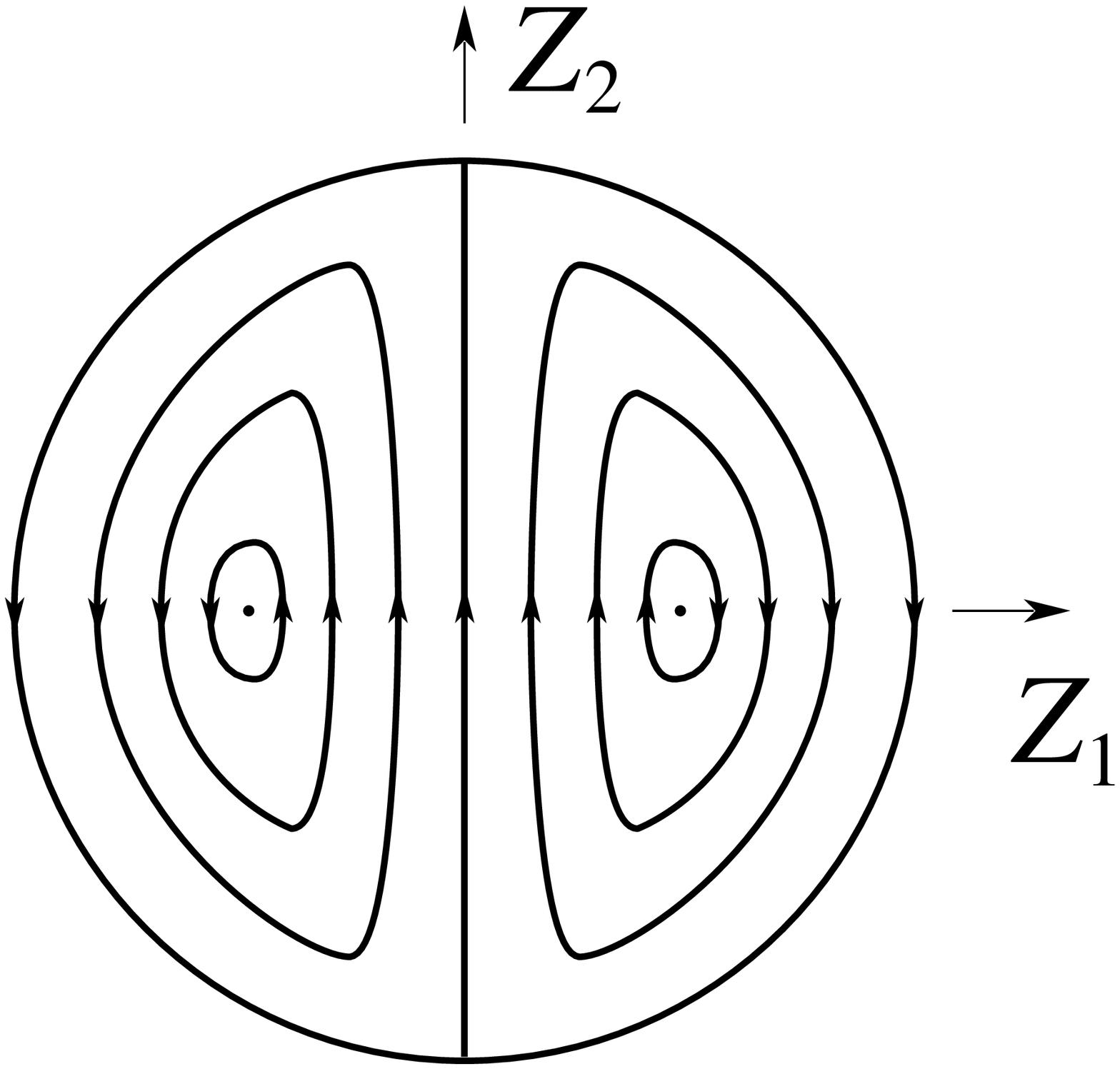}
      \vfill
\noindent
Figure 3. Phase portrait in the $Z$-plane for $J=0$.
The motion is confined within the circle $|Z|^2=\half N$.
The segment of the imaginary axis within this circle is
the homoclinic orbit.
      \vfill
\hrule height 1pt
\eject
\fi

\ifnum\Figsinline=0
\hrule height 1pt
    \vfill
    \hglue -10 true mm
    \epsfxsize=6.0 true in
\flushleft
\includegraphics[scale=0.8,angle=0]{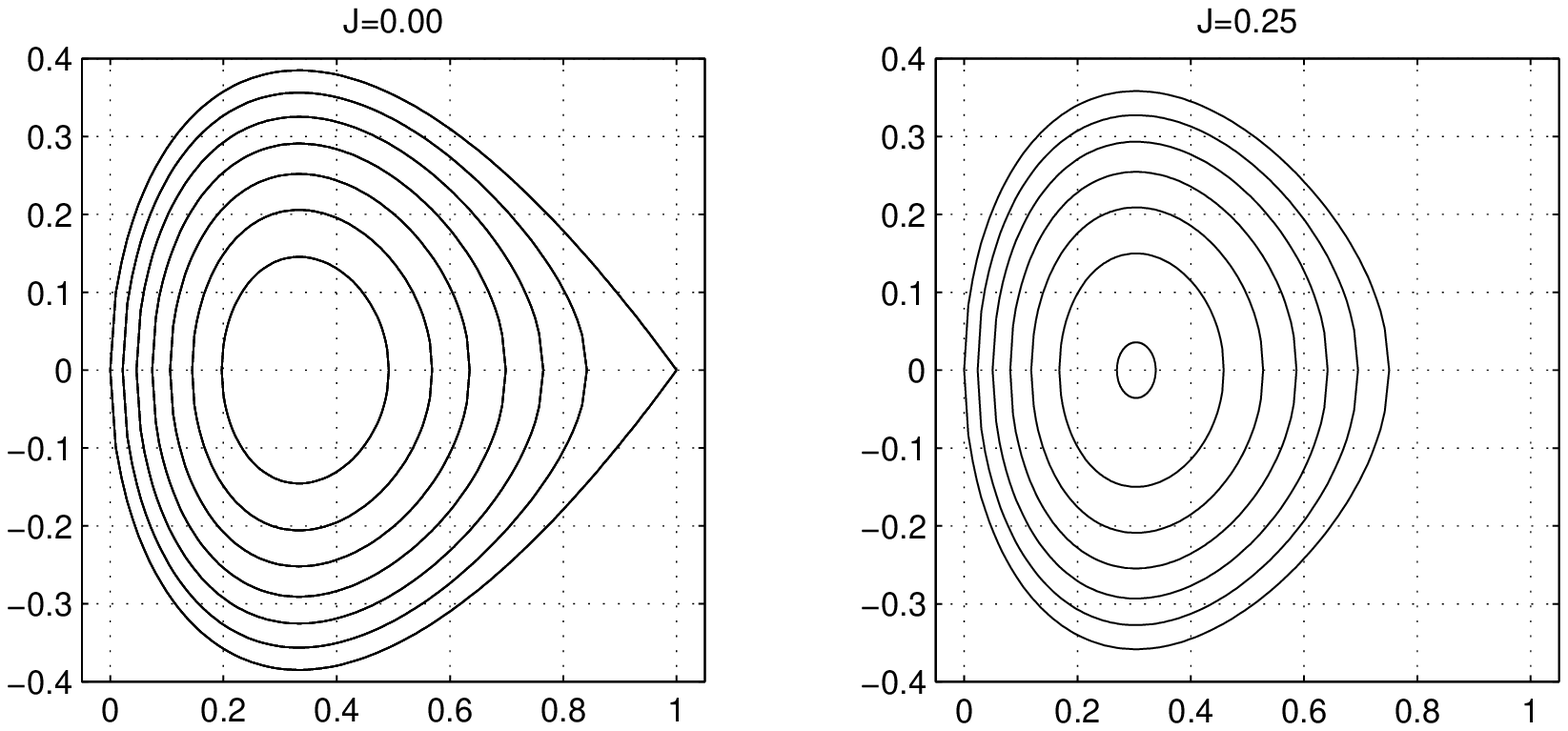}
    \vfill
\noindent
Figure 4.  Plots of  $\dot {\cal Q}$ versus ${\cal Q}$
for ${\cal J}=0$ and ${\cal J}=0.25$, for a range of values
\hfill\break\qquad
$E \in \{ -0.0635, -0.0529, -0.0423, -0.0317, -0.0212, -0.0106, 0 \}$.
    \vfill
\hrule height 1pt
\eject
\fi

\ifnum\Figsinline=0
\hrule height 1pt
    \vfill
    \hglue -15 true mm
    \epsfxsize=5.0 true in
\flushleft
\includegraphics[scale=0.8,angle=0]{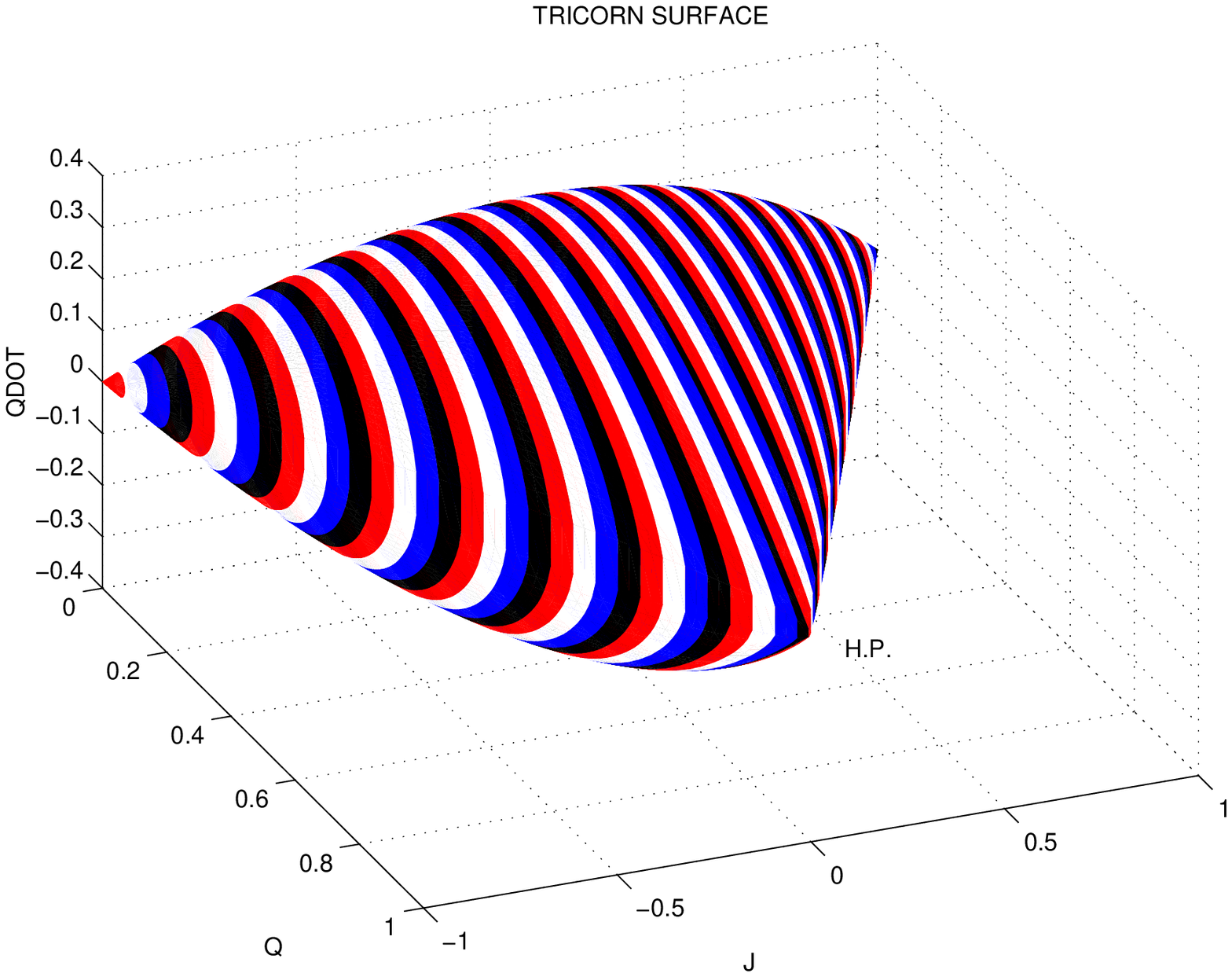}
    \vfill
\noindent
Figure 5.  Tricorn surface, upon which motion takes place when $H=0$.
The coordinates are ${\cal J}$, ${\cal Q}$, $\dot {\cal Q}$.
The motion takes place on the intersections of this surface with a plane
of constant $\cal J$ (such planes are indicated by the stripes).
This surface has three singular points.
The homoclinic point is marked H.P.
    \vfill
\hrule height 1pt
\eject
\fi

\ifnum\Figsinline=0
\hrule height 1pt
    \vfill
    \hglue -10 true mm
    \epsfxsize=5.0 true in
\flushleft
\includegraphics[scale=0.8,angle=0]{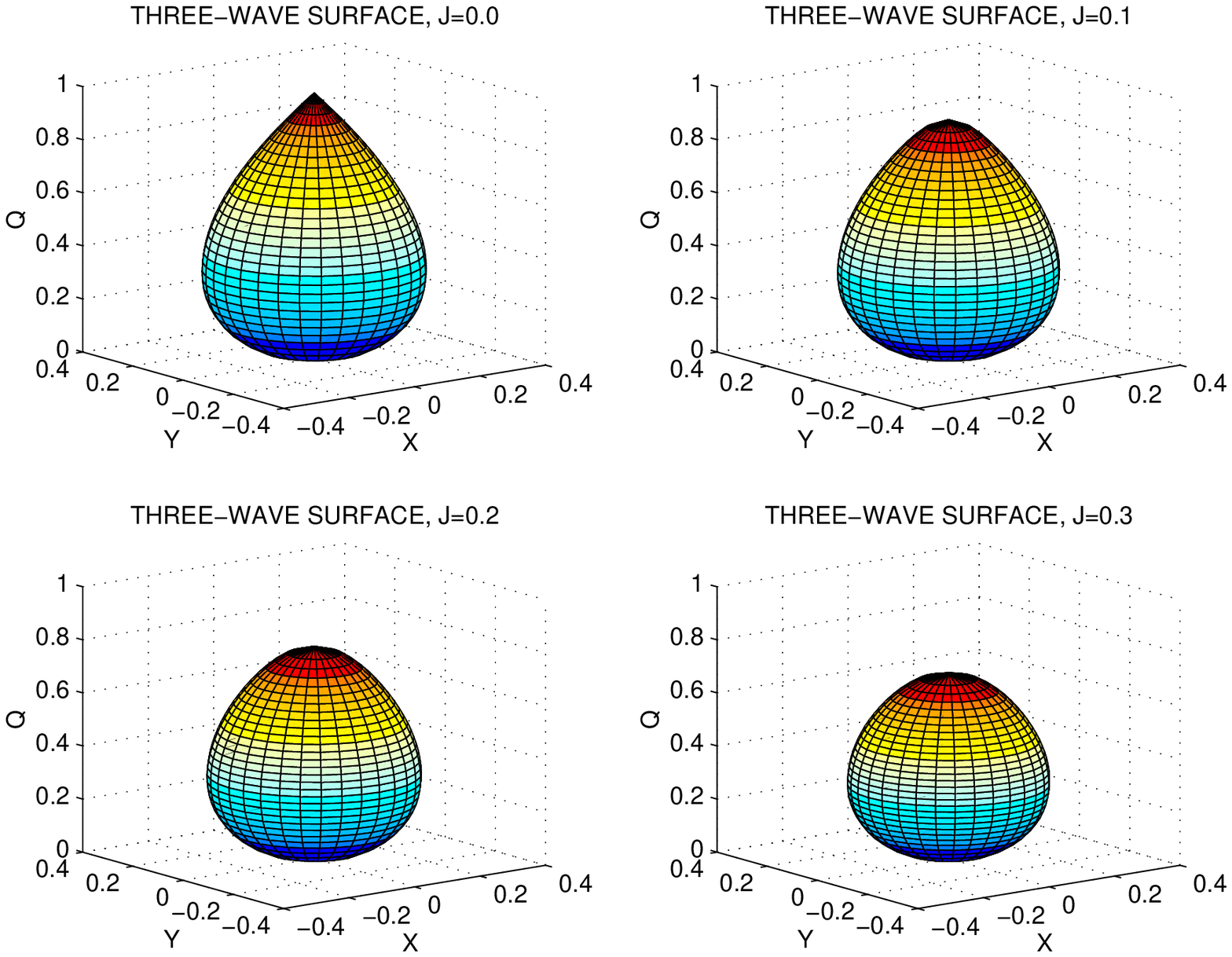}
    \vfill
\noindent
Figure 6. Surfaces of revolution about the ${\cal Q}$-axis, for
${\cal J}\in\{0.0,0.1,0.2,0.3\}$.
The radius for given ${\cal Q}$ is given
by the square-root of the cubic $-{\cal V}({\cal Q})$.
For given ${\cal J}$, the motion takes place on the intersection
of the corresponding surface with a plane of constant $X$.
    \vfill
\hrule height 1pt
\eject
\fi

\ifnum\Figsinline=0
\hrule height 1pt
    \vfill
    \epsfxsize=3.0 true in
\flushleft
\includegraphics[scale=0.4,angle=-90]{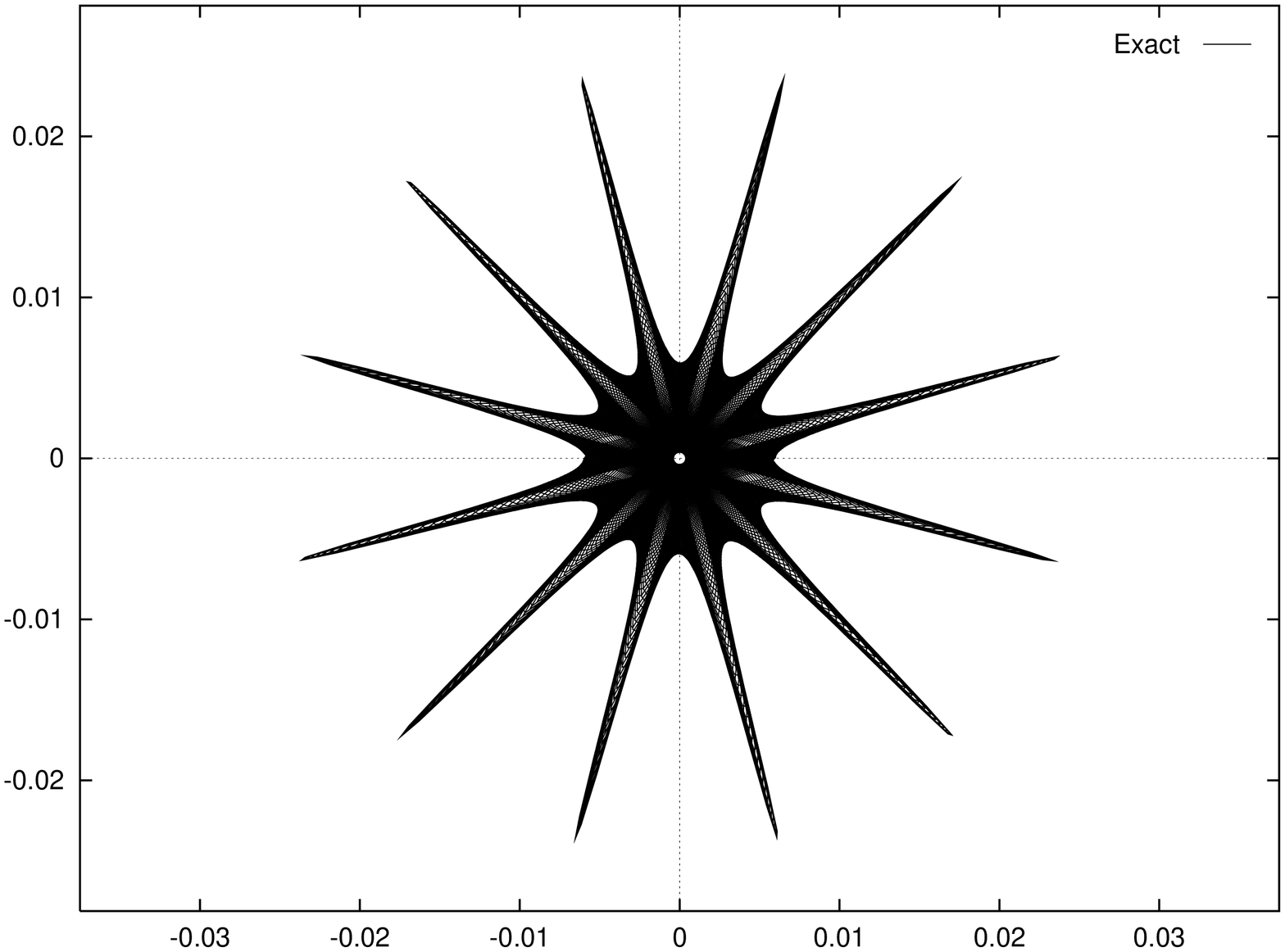}
    \hglue 1.8cm
    \rotr{201}
    \vskip 10 true mm
    \epsfxsize=3.0 true in
\flushleft
\includegraphics[scale=0.4,angle=-90]{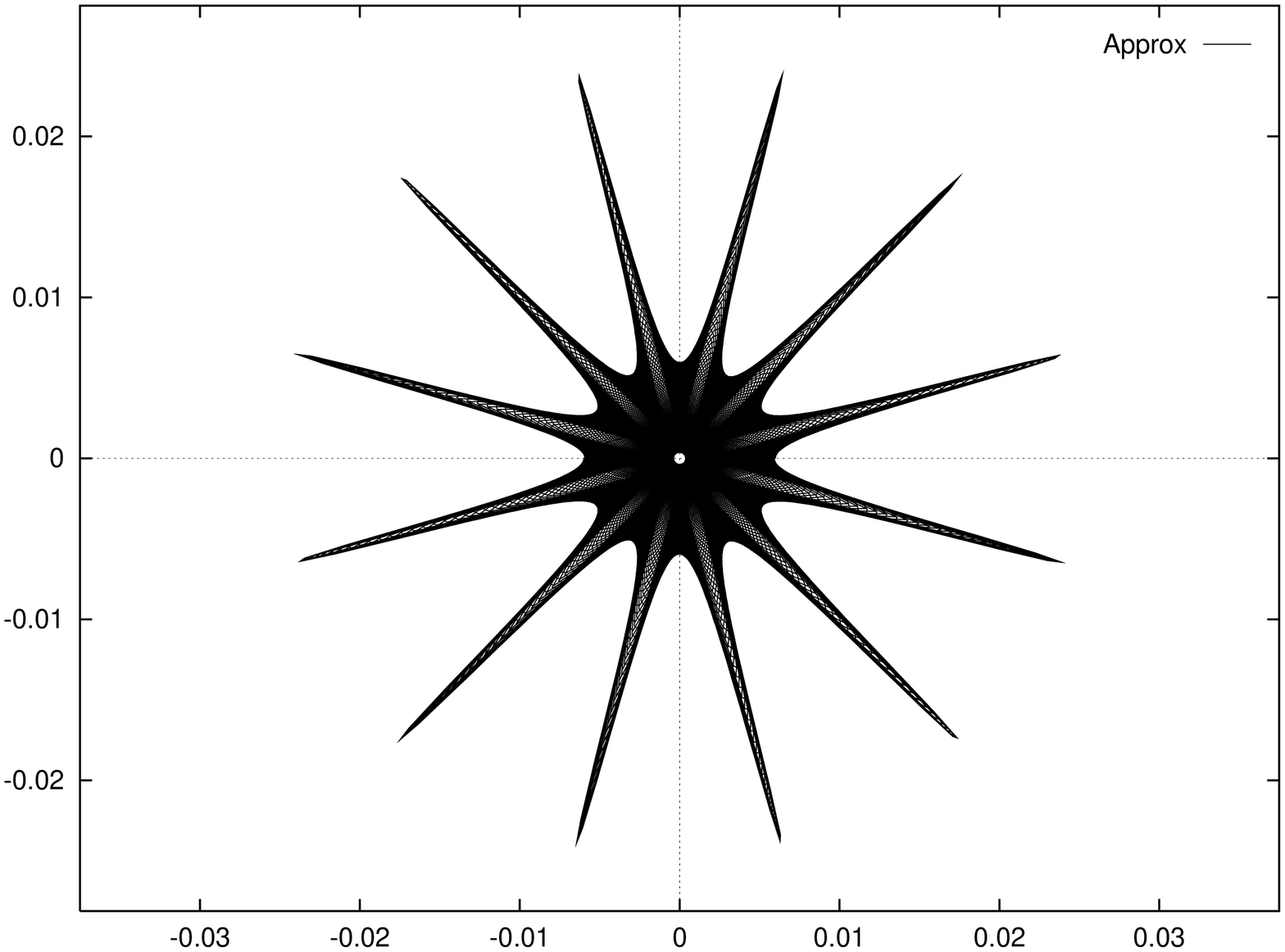}
    \hglue 1.8cm
    \rotr{202}
    \vfill
\noindent
Figure 7. Horizontal projection of the solution for an integration
of 1000 seconds. Top: solution of the `exact' equations.
Bottom: solution of the `approximate' equations.
    \vfill
\hrule height 1pt
\eject
\fi

\ifnum\Figsinline=0
\hrule height 1pt
    \vfill
    \epsfxsize=3.0 true in
\flushleft
\includegraphics[scale=0.4,angle=-90]{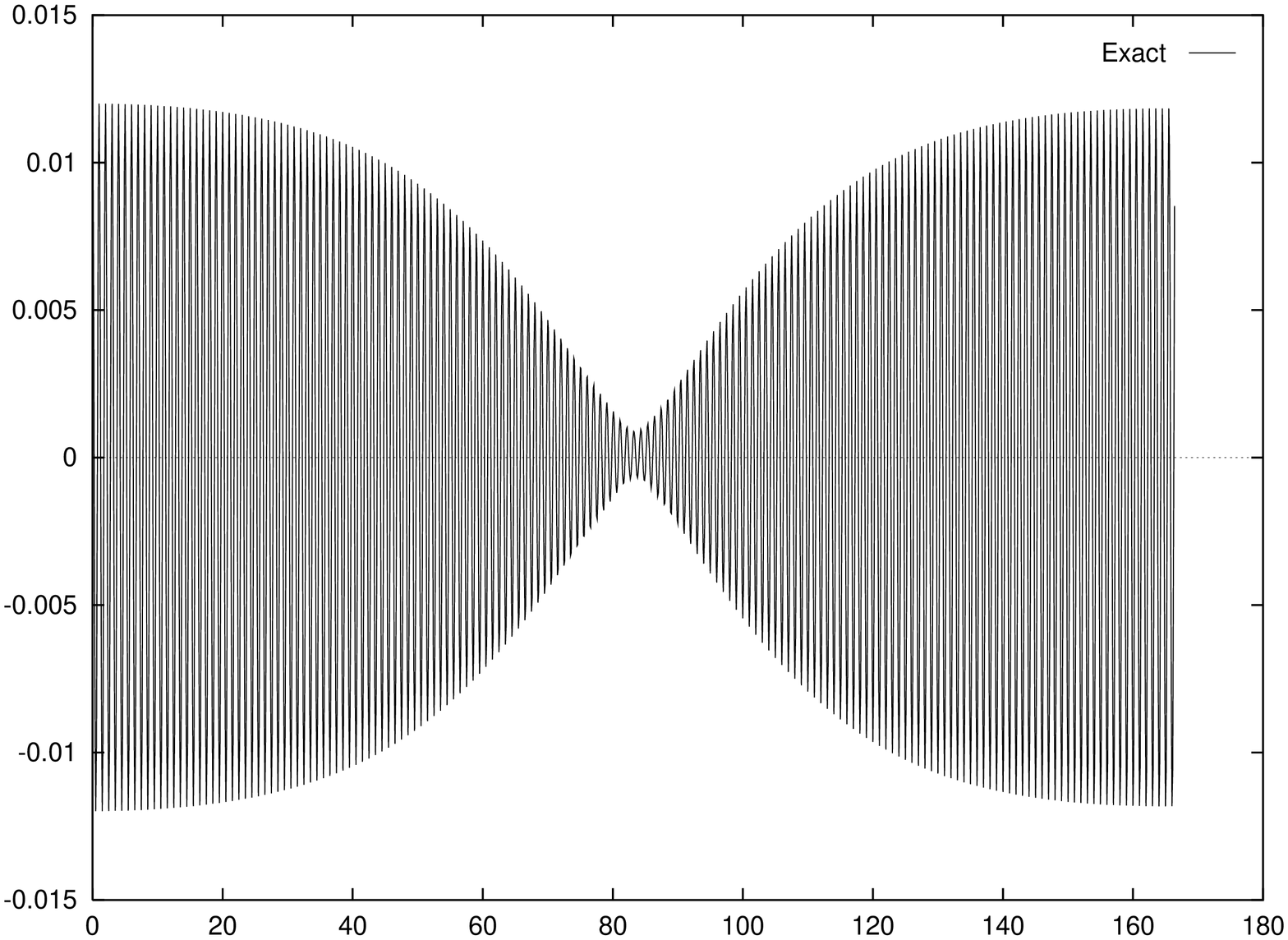}
    \hglue 1.8cm
    \rotr{203}
    \vskip 10 true mm
    \epsfxsize=3.0 true in
\flushleft
\includegraphics[scale=0.4,angle=-90]{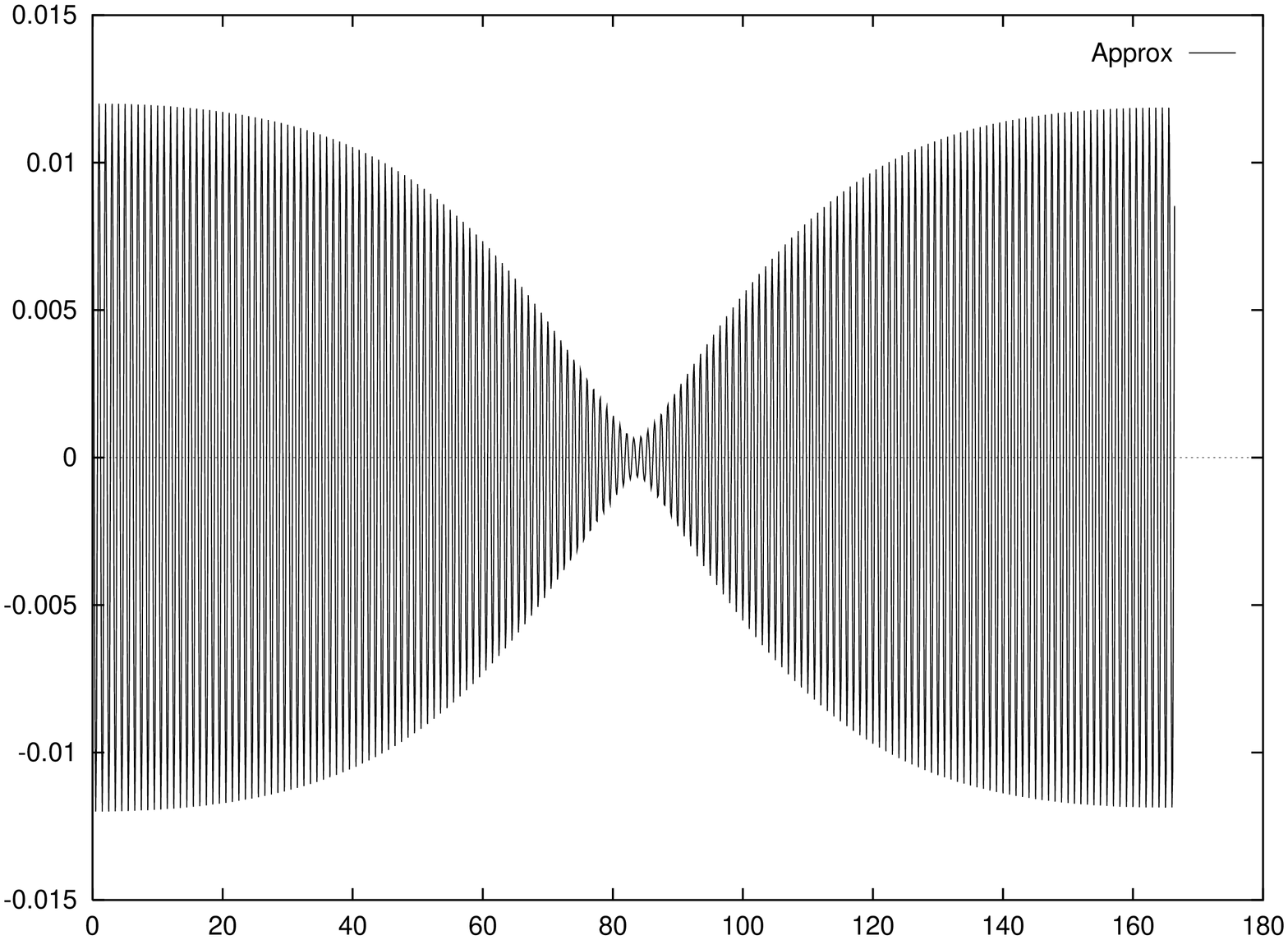}
    \hglue 1.8cm
    \rotr{204}
    \vfill
\noindent
Figure 8. Vertical amplitude of the solution for the
first modulation cycle (first 167 seconds).
Top: solution of the `exact' equations.
Bottom: solution of the `approximate' equations.
    \vfill
\hrule height 1pt
\eject
\fi

\ifnum\Figsinline=0
\hrule height 1pt
    \vfill
    \epsfxsize=3.0 true in
\flushleft
\includegraphics[scale=0.4,angle=-90]{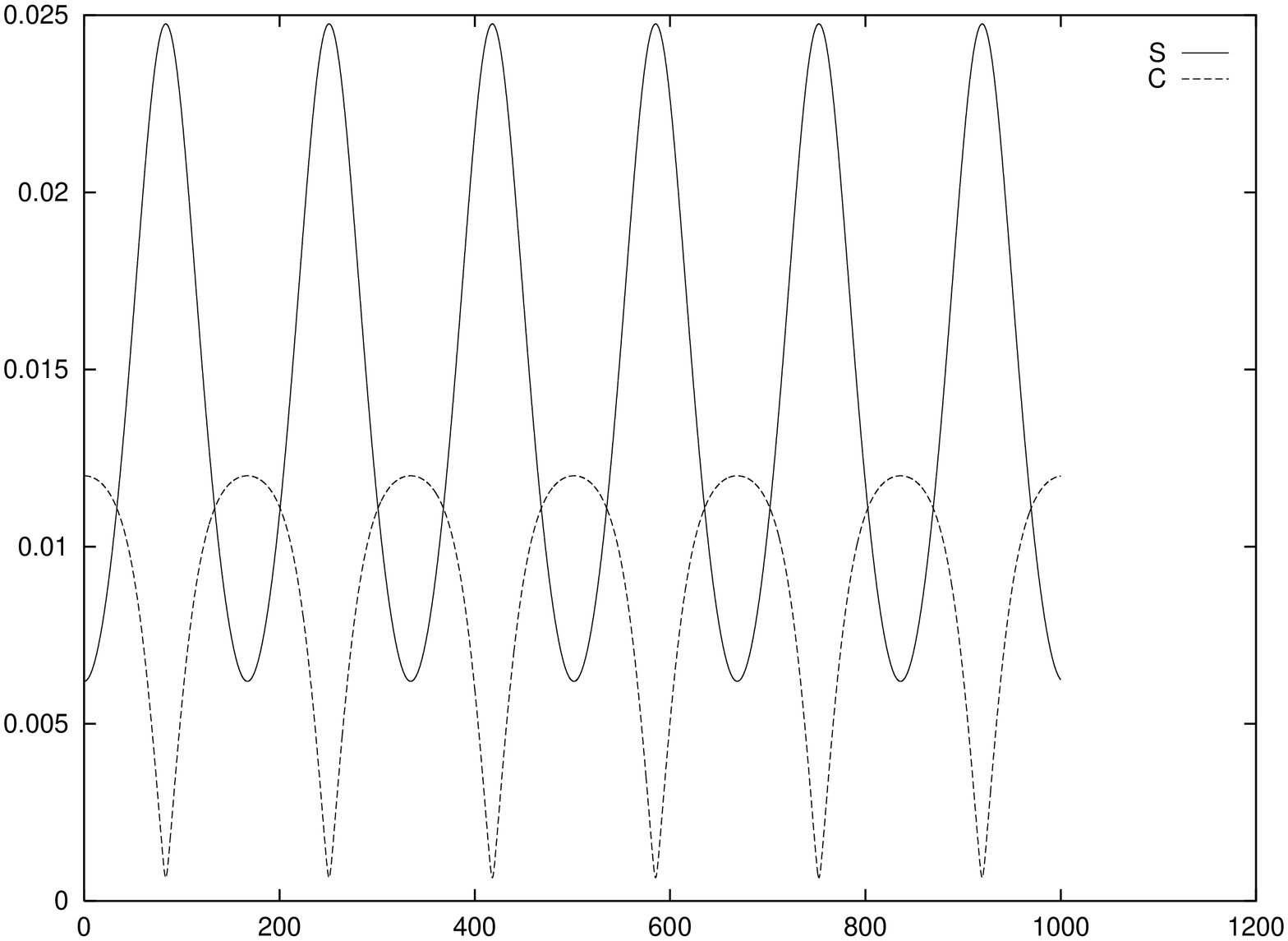}
    \hglue 1.0cm
    \rotr{205}
    \vglue 0.5cm
    \vskip 10 true mm
    \epsfxsize=3.0 true in
\flushleft
\includegraphics[scale=0.4,angle=-90]{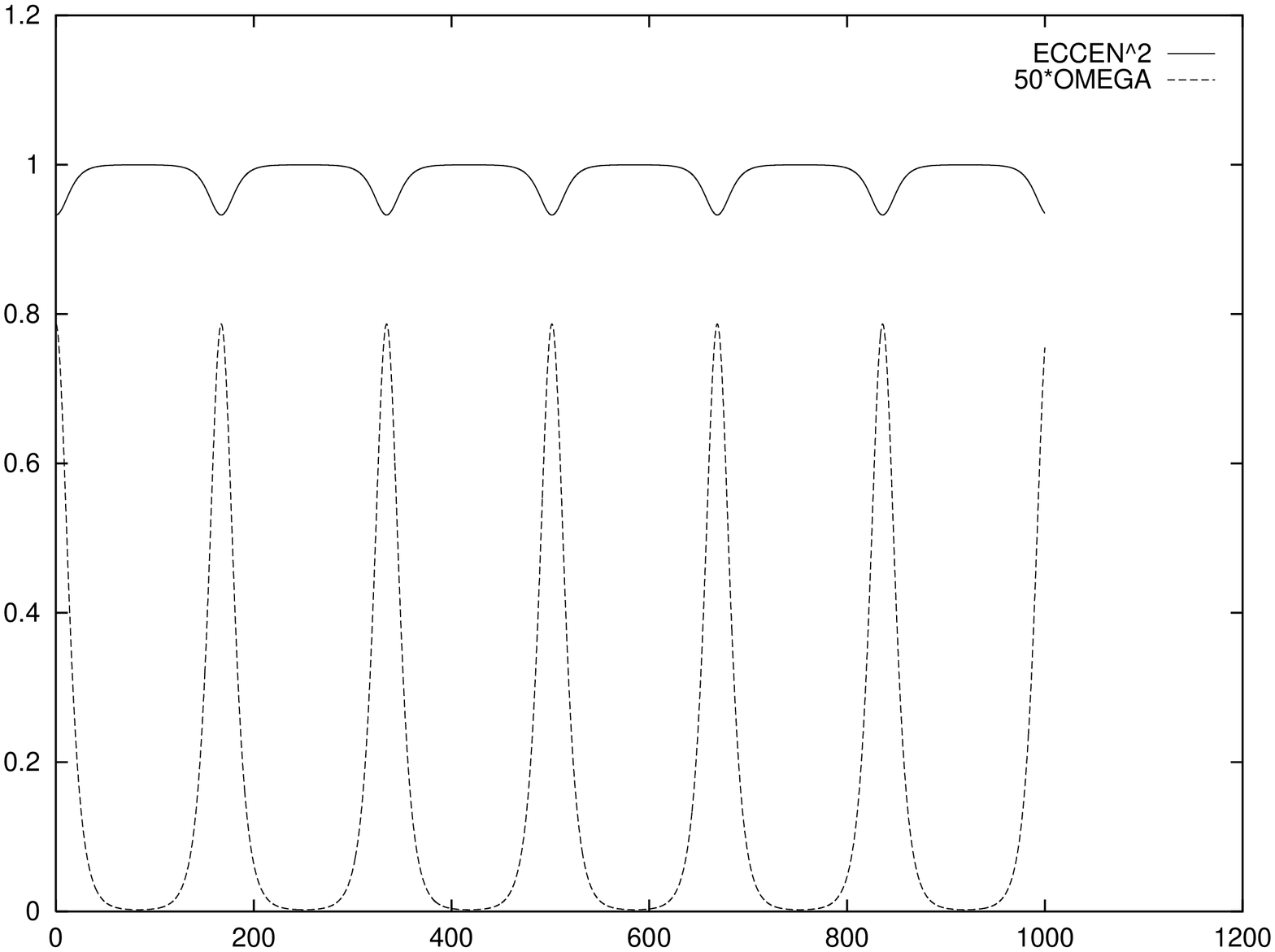}
    \hglue 1.0cm
    \rotr{206}
    \vfill
\noindent
Figure 9.  Top panel: Envelope amplitude of the approximate solution.
$S=\sqrt{|a|^2+|b|^2}$ (solid line) and $C=|c|$ (dashed line).
Bottom panel: Square of the eccentricity (solid line) and
angular velocity $\Omega$ (scaled by 50) of the instantaneous ellipse
(dashed line).
    \vfill
\hrule height 1pt
\eject
\fi

\ifnum\Figsinline=0
\hrule height 1pt
    \vfill
    \epsfxsize=3.0 true in
\flushleft
\includegraphics[scale=0.6,angle=-90]{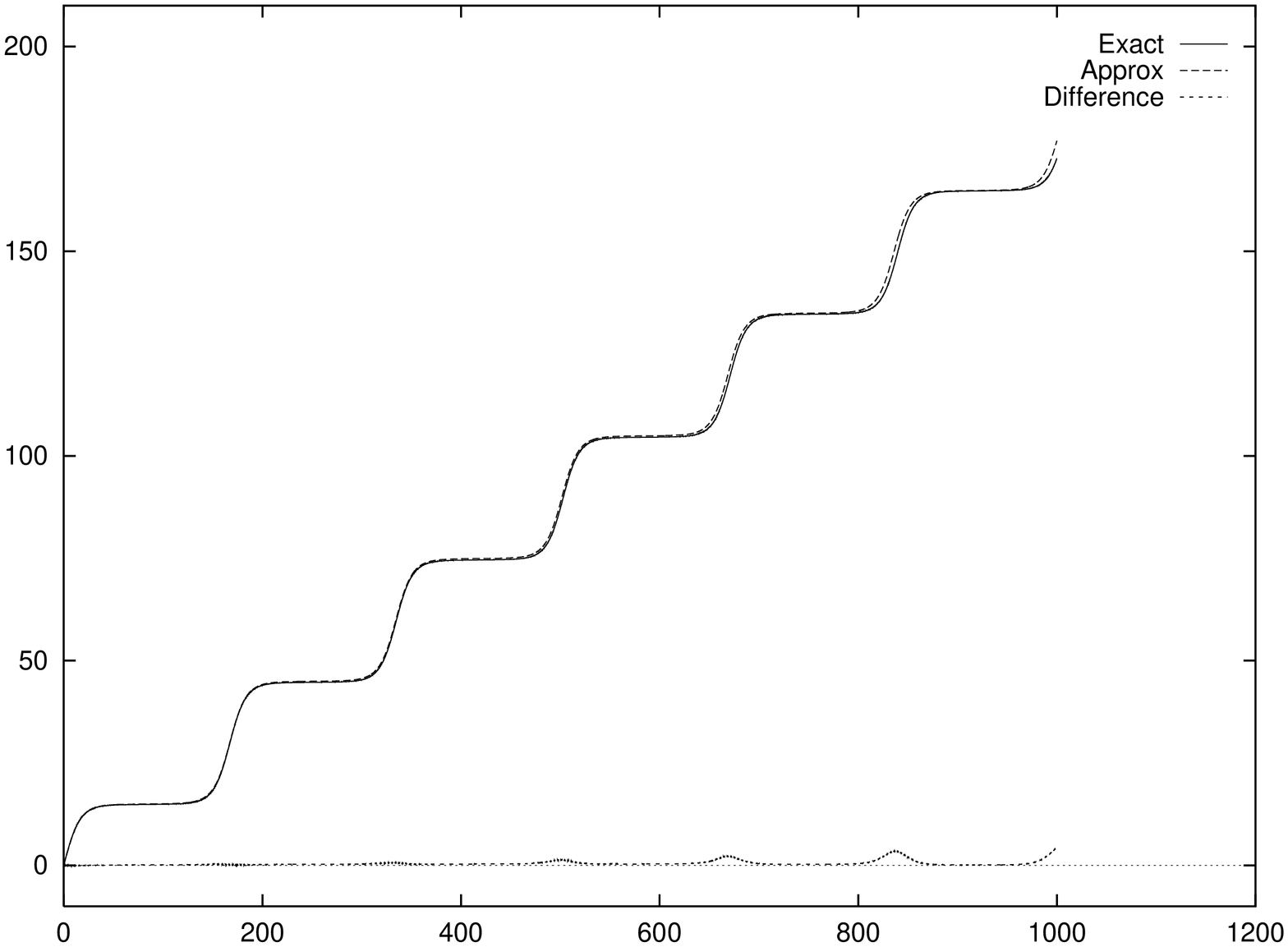}
    \rotr{205}
    \vfill
\noindent
Figure 10. Azimuth angle (in degrees) for the `exact' solution
($\vartheta$, solid line) and the `approximate' solution
($\theta$, dashed line).
The difference $\theta-\vartheta$ is plotted as a dotted line.
The azimuth $\Theta$ resulting from integration of (\ref{OmegaSpec})
(not plotted) is indistinguishable from the values $\theta$ of the
approximate solution.
    \vfill
\hrule height 1pt
\eject
\fi

\bye